\documentclass[notitlepage,prd,noshowpacs,noshowkeys,preprintnumbers,%
nofootinbib,11pt]{revtex4}
\pdfoutput=1 %necessary for pdf figures (pdflatex) to JHEP
\usepackage[latin1]{inputenc}
\usepackage{ae,aecompl}
\usepackage{amsmath,amssymb,mathrsfs}
\usepackage{graphicx}
\usepackage[usenames,dvipsnames]{color} % p usar nomes de cores
\usepackage{hyperref}
\hypersetup{
   colorlinks=true,       % false: boxed links; true: colored links
    linkcolor=Blue,          % color of internal links
    citecolor=Plum,        % color of links to bibliography
    filecolor=magenta,      % color of file links
    urlcolor=YellowOrange           % color of external links
}

\providecommand{\ZZ}{\mathbb{Z}}
\providecommand{\mss}[1]{\mbox{\scriptsize $#1$}}
\providecommand{\ml}[1]{\mbox{\large $#1$}}
\providecommand{\tp}{{\mss{\mathsf{T}}}}
\providecommand{\eq}[1]{\begin{equation} #1 \end{equation}}
\providecommand{\eqali}[1]{\begin{equation}\begin{aligned} #1
    \end{aligned}\end{equation}}
\providecommand{\eqarr}[1]{\begin{eqnarray} #1 \end{eqnarray}}
\providecommand{\mtrx}[1]{\begin{pmatrix} #1 \end{pmatrix}}
\DeclareMathOperator{\im}{\mathrm{Im}} % small space afte symbol
\DeclareMathOperator{\re}{\mathrm{Re}} % small space afte symbol
\DeclareMathOperator{\diag}{\mathrm{diag}} % small space after symbol
\usepackage{bbm}
\providecommand{\id}{{\mathbbm{1}}} %scalable
\providecommand{\bs}[1]{\boldsymbol{#1}}
\providecommand{\aver}[1]{\langle #1 \rangle}
\providecommand{\ums}[2][1]{\ml{\tfrac{#1}{#2}}} %small fraction in equations
\providecommand{\xlink}[1]
  {\small\href{http://arxiv.org/abs/#1}{arXiv:#1}}

\providecommand{\hx}{\hat{x}}
\providecommand{\bx}{\mathbf{x}}

\providecommand{\cp}{\mathsf{CP}}
\providecommand{\cL}{\mathtt{L}}
\providecommand{\cB}{\mathtt{B}}
\providecommand{\dcp}{\delta_{\cp}}

\providecommand{\om}{\omega}
\providecommand{\zcp}[1][2]{\ZZ^{\cp}_{#1}}

\providecommand{\lag}{\mathcal{L}}

\providecommand{\tphi}{\tilde{\phi}}
\providecommand{\eps}{\epsilon}
\providecommand{\veps}{\varepsilon}
\providecommand{\Umutau}{\mathrm{U(1)}_{\mu-\tau}}
\providecommand{\cpmutau}{\cp^{\mu\tau}}
\providecommand{\Rz}{R^{(0)}}
\providecommand{\Uz}{U^{(0)}}
\providecommand{\Mp}{M_{\rm pl}}
\providecommand{\lsym}{\lambda_\mathrm{sym}}

\providecommand{\tm}{\tilde{m}}
\providecommand{\hM}{\hat{M}}

\begin{document}
%%%%%%%%%%%%%%%%%%%%%%%%%%%%%%%%%%%%%%%%%%%%%%%%%
\preprint{UMD-PP-015-009}
\title{Implications of $\mu-\tau$ Flavored CP Symmetry of Leptons}
\author{R.~N.~Mohapatra$^{1}$}
\email{rmohapat@umd.edu}
\author{C.~C.~Nishi$^{1,2}$}
\email{celso.nishi@ufabc.edu.br}
\affiliation{\small $^1$ Maryland
Center for Fundamental Physics and Department of Physics,
University of Maryland, College Park, Maryland 20742, USA
}
\affiliation{\small $^2$
Universidade Federal do ABC - UFABC, 09.210-170,
Santo André, SP, Brasil
}

%\date{\today}
%%%%%%%%%%%%%%%%%%%%%%%%%%%%%%%%%%%%%%%%%%%%%%%%%
\begin{abstract}
We discuss gauge models incorporating $\mu-\tau$
flavored CP symmetry (called $\cpmutau$ in the text) in combination
with $L_\mu-L_\tau$ invariance to understand neutrino mixings and discuss their
phenomenological implications. We show that viable
leptogenesis in this setting requires that the lightest right-handed neutrino mass
must be between $10^9-10^{12}$ GeV and for effective two hierarchical right-handed
neutrinos, leptogenesis takes place
only in a narrower range of $5\times 10^{10}-10^{12}$ GeV.
A multi-Higgs  realization of this idea implies that there must be a pseudoscalar Higgs boson
with mass less than 300 GeV.
Generically, the vev alignment problem can be naturally avoided in our setting.
 \end{abstract}
%%%%%%%%%%%%%%%%%%%%%%%%%%%%%%%%%%%%%%%%%%%%%%%%%
% \pacs{12.60.Fr, 11.30.Er, 11.30.Fs, 11.30.Qc}
% 14.80.Ec 	Other neutral Higgs bosons
% 14.80.Fd 	Other charged Higgs bosons
% 11.30.Fs 	Global symmetries (e.g., baryon number, lepton number)
% 11.30.Qc 	Spontaneous and radiative symmetry breaking
% \keywords{Higgs, multi-Higgs models, 2HDM, global symmetry, custodial symmetry,
% symmetry breaking}
%\twocolumn
\maketitle
% \tableofcontents
%%%%%%%%%%%%%%%%%%%%%%%%%%%%%%%%%%%%%%%%%%%%%%%%%
\section{Introduction}
\label{sec:intro}

Phenomenal success of experimental research in neutrino physics in the last two decades
have led not only to unequivocally establishing that neutrinos have mass but also
to an almost complete determination of flavor mixings between the different lepton
generations.
The missing parts are: (i) the Dirac CP phase, (ii) neutrino
mass hierarchy and (iii) a knowledge of whether neutrinos are Majorana or Dirac
fermions. Assuming that there are no extra sterile neutrinos,
the discovery of the CP phase for neutrinos  would put flavor information
on leptons on the same footing as quarks. If neutrinos are Majorana fermions, there 
would be two more phases present in the flavor space and for complete information, 
one will need information on them. The latest global fits \cite{valle:fit,GG:fit} of 
neutrino parameters point to a preference for a negative value for the Dirac CP 
phase, $-\pi<\dcp<0$.
A key focus of experimental research in neutrino physics at
the moment is therefore to determine the Dirac CP phase in addition to answering the question of whether neutrinos are Dirac or Majorana particles and their mass hierarchy.
An additional motivation to determine  the Dirac CP phase comes from its possible connection to understanding the
origin of matter and anti-matter asymmetry in the universe via
leptogenesis~\cite{lepto_rev}. While it is well known that non-observation of a
non-zero Dirac CP phase does not preclude leptogenesis, its observation can 
nonetheless provide important insight into the latter~\cite{pascoli}.

On the theory front, understanding of the lepton mixing angles $\theta_{ij}$ has been one of the two
major driving forces of much of the research in this field, the other being to probe the scale of neutrino masses. In the former
case, symmetries have been used as a main tool, motivated by the observation that mixing angles $\theta_{23}\sim \frac{\pi}{4}$
and ${\rm sin}\theta_{12}\sim \frac{1}{\sqrt{3}}$, suggesting their possible group
theoretic origin~\cite{review}. Among the very first symmetries studied for
neutrinos is the $\mu$-$\tau$ exchange symmetry~\cite{mutauexc}, which not only
predicted maximal $\theta_{23}$ but also that $\theta_{13}=0$. Many other symmetries
such as $S_4$, $A_4$, $\Delta(3n^2)$, etc., were considered later on. The so-called
tri-bi-maximal (TBM) mixing pattern~\cite{hps} which embodied all these three features, i.e.,
$\theta_{23}\sim \frac{\pi}{4}$, ${\rm sin}\theta_{12}\sim \frac{1}{\sqrt{3}}$ as
well as $\theta_{13}=0$, together with the symmetry techniques to obtain this
pattern, gave a big boost to this approach.
Discovery of a non-zero and large value for $\theta_{13}$~\cite{expt} was a turning point in
this research since it ruled out the tri-bi-maximal mixing pattern. Since then, many
attempts have been made to combine flavor symmetries with CP transformation to
accommodate a non-zero $\theta_{13}$ while trying to predict the Dirac CP
phase~\cite{GL,cpmutau:others,fcp,fcp:others}, sometimes without imposing CP
explicitly\,\cite{babu,cp:real}.

In this paper, we pursue this line of research and consider a simple approach based
on a generalized definition of CP transformation that mixes it with $\mu$-$\tau$
exchange (called $\cpmutau$ from now on)~\cite{GL}. This symmetry is known to accommodate a
non-zero $\theta_{13}$ while at the same time  predicting a Dirac CP phase $\delta
\sim \pm90^0$~\cite{GL,babu} if the charged lepton mass matrices are taken diagonal.
There are also models where one has deviations from the exact $\cpmutau$
limit~\cite{deviation}.
A key challenge to building such models has been that in the $\cpmutau$ symmetry
limit, the muon and tau lepton Yukawa couplings are degenerate, leading to
same masses.
In Ref.\,\cite{GL}, explicit soft breaking terms were
introduced to generate the $\mu\tau$ mass splitting. Another uncomfortable feature of these models has been its apparent
inability to explain the origin of matter via leptogenesis~\cite{GL}. We address both these issues in this paper.
Our goal is to present a model where starting with a high scale symmetry, we find a
low energy effective theory where the neutrino sector maintains exact $\cpmutau$
symmetry whereas in the charged lepton sector, the symmetry is spontaneously broken so as to
allow the muon and tau masses to be different. We give two examples: one with an extended Higgs sector
and another with an extension involving heavy vector like fermions. The former has interesting
implications for Higgs physics that we discuss below. We also show that there exists a limited range of seesaw scales
where successful leptogenesis can take place, when lepton flavor effects are taken into account.

As a part of this investigation, we also identify the combination of family lepton numbers $\cL_\mu-\cL_\tau$~\cite{werner} (which
we denote as $\Umutau$) as the largest natural abelian symmetry that can be imposed
in conjunction with $\cpmutau$, thus providing the simplest example of combining an
abelian symmetry with CP, yet with predictive CP violation at low energies.
We arrive then at a natural setting where $G_l=\Umutau$ can be the residual
symmetry of the charged lepton sector (ensuring diagonal mass matrix) and
$G_\nu=\zcp$, generated by $\cpmutau$, is the residual symmetry of the
neutrino sector.
Because of the properties of $\Umutau$ and $\cpmutau$, these residual symmetries can
be maintained separately in each sector without perturbing interactions in the
scalar potential, thus avoiding the vev alignment problem of flavor symmetry models
with larger nonabelian groups.

New results of the paper are:
(i) construction of a model with natural residual symmetries $G_l$ and $G_\nu$ but
without soft breaking of  $\cpmutau$;  (ii) discussion of how one can implement successful leptogenesis in these models and constraints imposed by
it on the seesaw scale and (iii) implications for neutrino-less double beta decay and
Higgs physics.

This paper is organized as follows: in sec.\,\ref{sec:cpmutau}, we review the
consequences of $\cpmutau$ on the neutrino mass matrix and PMNS.
Sections \ref{sec:nuless} and \ref{sec:lepto} present general consequences of
$\cpmutau$ symmetry on neutrino-less double beta decay and leptogenesis.
In sec.\,\ref{sec:sym}, we introduce the generalized CP like
symmetries and show how  $\cpmutau$ symmetry emerges as the
trivial automorphism of gauged $U(1)_{\mu-\tau}$ symmetry.
We then present a multi-Higgs implementation of the symmetry in sec.\,\ref{sec:model},
together with some phenomenological implications.
Our paper is summarized in sec.\,\ref{sec:summary}.
The appendices contain the proof of the uniqueness of $\cpmutau$,
the $\cpmutau$ symmetry in the real basis
and another realization of the idea where $G_l\times G_\nu$ is exact at
high energies, which uses heavy vector like fermions instead of extra weak scale
Higgs doublets.

%%%%%%%%%%%%%%%%%%%%%%%%%%%%%%%%%%%%%%%%%%%
\section{Maximal $\theta_{23}$ and Dirac CP phase from $\cpmutau$: a review}
\label{sec:cpmutau}

The latest global fits \cite{valle:fit,GG:fit} of neutrino parameters still allows
maximal atmospheric angle $\theta_{23}=45^\circ$ within 2$\sigma$ and also
point to a preference for negative values for the Dirac CP phase,
$-180^\circ<\dcp<0$.
It was pointed out in \cite{GL} that maximal $\theta_{23}$ and maximal $\dcp$, i.e.,
\eq{
  \label{mutau-r:pred}
\theta_{23}=\pi/4 \text{~~and~~} \dcp=\pm \pi/2\,,
}
follow from  the neutrino mass matrix invariant under $\cpmutau$ symmetry.
In the flavor basis (fixed by some $G_l$), it corresponds to the relation:
\eq{
  \label{mutau-r}
X^\tp M_\nu X =M_\nu^*\,,
}
where
\eq{
X=\mtrx{1&0&0\cr0&0&1\cr0&1&0}\,.
}
Clearly, this symmetry can be implemented in the neutrino sector as the composition
of $\mu\tau$ interchange symmetry with CP conjugation.
We will show a simple and natural setting where this symmetry survives in the
neutrino sector but is broken in the charged lepton sector.

Let us review some aspects of $\cpmutau$.
First, the symmetry \eqref{mutau-r} implies a neutrino mass matrix of the form~ \cite{babu,GL}
\eq{
  \label{mutau-r:form}
M_\nu=\left(
\begin{array}{ccc}
 a & d & d^* \\
 d & c & b \\
 d^* & b & c^* \\
\end{array}
\right)\,,
}
where $a,b$ are real whereas $c,d$ are complex a priori.
It is necessary that both $c\neq 0$, $d\neq 0$, and $\im(d^2c^*)\neq 0$, to ensure
$\theta_{13}\neq 0$\,\cite[b]{GL} because a rephasing transformation can
turn $M_\nu$ to a matrix invariant under the simpler (unitary) $\mu\tau$
interchange symmetry.

One can show that a matrix of the form \eqref{mutau-r:form} can be always
diagonalized by a matrix of the form\,\cite{GL}
\eq{
  \label{U0}
U_0=\mtrx{u_1&u_2&u_3\cr w_1&w_2&w_3\cr w_1^*&w_2^*&w_3^*}
\,,
}
where $u_i$ are real and conventionally positive.
Application of complex conjugation on $M_\nu$ and $U_0$ shows that the
diagonalization of \eqref{mutau-r:form},
\eq{
  \label{diag:U0}
U_0^\tp M_\nu U_0=\diag(\pm m_i)\,,
}
already leads to real diagonal entries, so that the Majorana phases are trivial,
i.e., either $1$ or $i$.
Therefore, we can write for the complete diagonalization matrix,
\eq{
  \label{U:cpmutau}
U_\nu=U_\nu^{(0)}K_\nu\,,
}
where $U_\nu^{(0)}$ has the form \eqref{U0} and $K_\nu$ is diagonal and
contains the Majorana phases $(K_\nu)_{ii}=1\text{ or }i$.
We denote the different possibilities by
\eq{
\text{diagonal of $K_\nu^2$}  \sim (+++),(-++),(+-+) \text{~or~}(++-)\,,
}
which correspond to the CP parities of $\nu_{iL}$ assuming $\cpmutau$.

It is easy to see that $U_0$ obeys
\eq{
|(U_0)_{\mu j}|=|(U_0)_{\tau j}|\,,\quad\text{for $j=1,2,3$}.
}
The equality for $j=3$ signals maximal $\theta_{23}$. The equality for $j=1,2$,
easily seen in the standard parametrization, leads to\,\cite{GL}
\eq{
\sin\theta_{13}\sin\dcp=0\,.
}
This signals maximal $\dcp$ since $\theta_{13}\neq 0$.

\section{Neutrino-less double beta decay in theories with $\cpmutau$}
\label{sec:nuless}

For Majorana neutrinos, there is a nonzero probability of neutrino-less double beta
decay to occur.
The rate depends on the square of the modulus of
\eq{
m_{ee}\equiv \sum_i m_iU_{ei}^2\,.
}
In general, this quantity depends on the Dirac CP phase (depending on the
convention) and Majorana CP phases.
For the theory invariant under $G_\nu=\zcp$ and $G_l\subset \Umutau$,
$\dcp=\pm\pi/2$, only a discrete choice of possibilities for the Majorana
phases remain.
We obtain
\eq{
m_{ee}= \sum_i m'_i{U_{ei}^{(0)}}^2\,,
}
where $U_{ei}^{(0)}$ are real positive quantities fixed by
$\theta_{12},\theta_{13}$, cf. \eqref{U:cpmutau},
and $m_i'=\pm m_i$ are the light neutrino masses with its CP parities.

\begin{figure}[h]
\centering
\includegraphics[scale=0.45,angle=0]{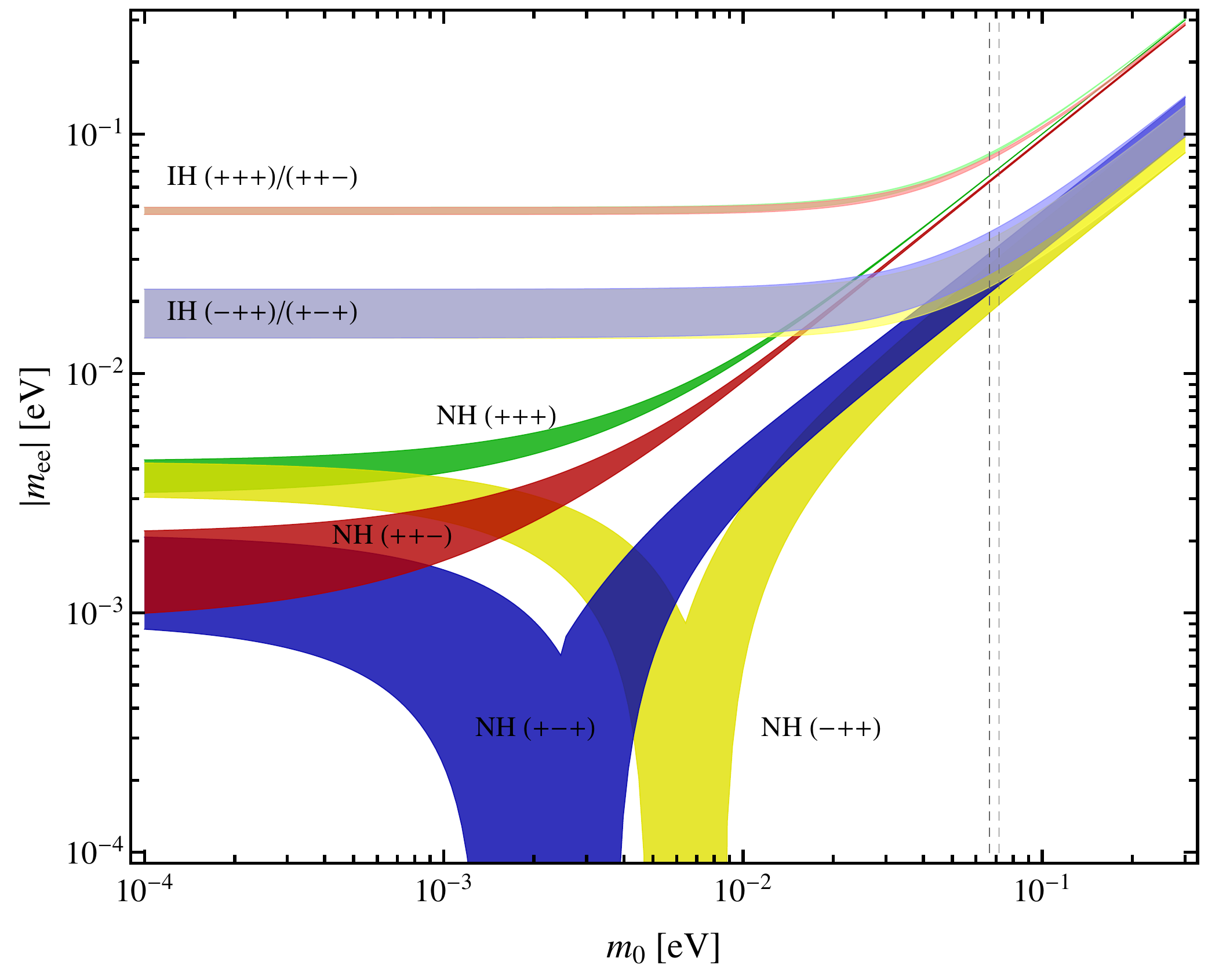}
\caption{
$|m_{ee}|$ as a function of the lightest mass $m_0$ ($m_1$ for NH and $m_3$ for IH)
for CP parities $K_{\nu ii}^2$ of the light neutrinos $\nu_{iL}$: $(+++)$ (green),
$(-++)$ (yellow),
$(+-+)$ (blue) and $(++-)$ (red).
Darker colors denotes NH and lighter colors denotes IH. For the latter, light
blue and yellow (light red and green) are largely overlapped.
We use the 3-$\sigma$ allowed ranges for $\Delta m_{21}^2,\Delta m^2_{31}$,
$\theta_{12},\theta_{13}$ of Ref.\,\cite{GG:fit}.
The vertical dashed lines shows the current bound coming from the
cosmological data on $\sum m_i$; cf. \eqref{bound:cosmo}.
}
\label{fig:pred:mee}
\end{figure}

In Fig.\,\ref{fig:pred:mee} we show the discrete possibilities for $|m_{ee}|$ as a
function of the lightest neutrino mass $m_0$ ($m_1$ for NH and $m_3$
for IH). We vary $\Delta m_{21}^2$, $\Delta m^2_{31}$,
$\theta_{12},\theta_{13}$ within their 3-$\sigma$ allowed values\,\cite{GG:fit}
($\theta_{23}=\pi/4$ is fixed from symmetry).
We can see that some CP parities can be distinguished if independent information on
the mass hierarchy and sufficiently precise information of the absolute mass scale
is known. Specially for IH, we can distinguish between $(+++)/(++-)$ CP parities
for $\nu_L$ and $(-++)/(+-+)$.
For NH, some cases can be distinguished for some ranges of the absolute mass scale.
For example the $\tilde{S}_4$ ($A_4\rtimes \zcp$) model of
Ref.\,\cite[b]{fcp} lies in the lower (NH) yellow $(-++)$ band.
With enough precision, even in the quasi-degenerate spectrum we can distinguish
between $(+++)/(++-)$ and $(-++)/(+-+)$ CP parities.
Notice that some bands would completely overlap in the $m_0\to 0$ limit.
Regions similar to the ones we show here can be seen, in the
general phenomenological analysis of Ref.\,\cite{nuless} (see
its Fig.\,2 with dashed curves denoted as $(\pm\pm)$), but without the underlying
symmetry discussion.
Note that this predictions for neutrinoless double beta decay
is the same as for the strictly CP conserving case
at low energies but in our case the Dirac CP phase is
maximal instead of being $0$ or $\pi$, a fact that can
be distinguished in future oscillation experiments.

Also shown in Fig.\,\ref{fig:pred:mee} are the cosmological bounds for $m_0$,
\eqali{
  \label{bound:cosmo}
\text{NH}:&& m_0=m_1&<0.0716\,\text{eV}\,,\cr
\text{IH}:&& m_0=m_3&<0.0665\,\text{eV}\,.
}
These values are obtained from the cosmological bound of $\sum m_i<0.23$ at 95\%
C.L. reported by the Planck collaboration\,\cite{planck}
when 3-$\sigma$ range of $\Delta m^2_{21}$ and $\Delta m^2_{31}$ are considered.

%%%%%%%%%%%%%%%%%%%%%%%%%%%%%%%%%%%%%%%%%%%
\section{Leptogenesis}
\label{sec:lepto}

Neutrino mass mechanisms are widely considered to have a connection to the origin of matter via leptogenesis~\cite{nardi}.
In this section, we discuss this in the class of models we are discussing here.
The first consideration of leptogenesis with $\cpmutau$ symmetry was made in
\cite[b]{GL}. The authors concluded that leptogenesis is not possible because
$\cpmutau$ invariance of the neutrino sector ensured that all elements
$(\lambda\lambda^\dag)^2_{ij}$ were real leading to vanishing CP asymmetry, with
$\lambda$ being the $N_R$ Yukawa coupling in the basis where the RHNs are mass
eigenstates.
Such a conclusion, however, is only valid for the case where the heavy
neutrinos are hierarchical and charged lepton flavor effects are unimportant
(the so-called one-flavor approximation), i.e., for $T\sim M_{1}\gtrsim 10^{12}\rm
GeV$, where $M_1$ is the mass of the lightest right-handed neutrino.
Below that temperature, the tau lepton enters into thermal equilibrium due to its
Yukawa interaction with $\tau_R$ and flavor effects must be
considered (the so called flavored leptogenesis\,\cite{flavored}).
We will see that successful leptogenesis is possible even with $\cpmutau$ symmetry
in the intermediate range $10^9\text{GeV}\lesssim M_{1}\lesssim 10^{12}\rm GeV$ if
flavor effects are taken into account.
Therefore, we do not need small $\cpmutau$ breaking for successful leptogenesis as
in Ref.\,\cite[b]{deviation}.
Surprisingly, $\cpmutau$ symmetry seems to preclude successful leptogenesis for
$M_{1}\lesssim 10^9\rm GeV$ for hierarchical heavy right-handed neutrinos because both
$\tau$ and $\mu$ flavors are in thermal equilibrium; see
Sec.\,\ref{sec:hierachical}.
This result holds even if the resonant enhancement of CP asymmetries due to
quasi-degenerate heavy right-handed neutrinos are considered; see
Sec.\,\ref{sec:resonant}.

To prove our assertion, let us first review the consequences of $\cpmutau$ on the quantities relevant for
leptogenesis.
It is clear from the form of $U_\nu$ in \eqref{U:cpmutau} that $\cpmutau$ implies
the CP property
\eq{
\label{prop:U}
XU_\nu^*=U_\nu K_\nu^2 ~~\text{ or }~~
U_\nu^*=X^\dag U_\nu K_\nu^2\,.
}
This can be also generically inferred from the relation \eqref{mutau-r}.
As can be checked explicitly in the CP-basis, $K_\nu^2$ corresponds to the CP
parities of $\nu_{iL}$ considering $\cpmutau$ is conserved in the neutrino sector.
A similar relation is also valid for $U_R$, the matrix that diagonalizes $M_R$:
\eq{
U_R^*=XU_RK_R^2~~\text{and}~~U_R=U_R^{(0)}K_R\,.
}
Note that the previous relation assumes $M_R$ is in the symmetry
basis.
We also assume the charged lepton mass matrix (squared) is diagonal (flavor basis)
so that the PMNS matrix is $U=U_\nu$.

Let us write the type-I seesaw Lagrangian in the form
\eqali{
  \label{basis:lepto}
-\lag&= y_{\alpha}\bar{L}_\alpha H l_{\alpha R}
  +\bar{N}_{iR}\lambda_{i\alpha}\tilde{H}^\dag L_\alpha
  +M_i\bar{N}_{iR}N_{iR}^c
\,,
}
where the sum of repeated indices is implicit.
In this basis, the CP asymmetries depend only on $\lambda$ and the heavy masses
$M_i$.

In the symmetry basis, $\lsym$ obeys
\eq{
X^\dag\lsym X=\lsym^*\,.
}
In the basis of \eqref{basis:lepto}, we have
\eq{
\lambda=U_R^\dag\lsym\,,
}
and it obeys
\eq{
  \label{prop:lambda}
\lambda^*=K_R^2\lambda X\,.
}

\subsection{Hierarchical heavy neutrinos}
\label{sec:hierachical}

We can see the consequences of $\cpmutau$ on leptogenesis for the case where the
right-handed neutrinos $N_{iR}$ have hierarchical masses and only the decay of
lightest state $N_1$ is relevant for leptogenesis.
Our discussion, however, apply also to cases where the hierarchy is mild.
In our notation, the flavored CP asymmetries for the decay $N_1\to
l_\alpha+\phi$, $\alpha=e,\mu,\tau$, read (see e.g. \cite{nardi})
\eqali{
  \label{eps:lambda}
\eps_\alpha&=\frac{1}{8\pi(\lambda\lambda^\dag)_{11}}
\sum_{j\neq 1}\bigg\{
\im\big[(\lambda\lambda^\dag)_{j1}\lambda_{j\alpha}\lambda^*_{1\alpha}\big]g(x_j)
\cr
&\quad +\
\im\big[(\lambda\lambda^\dag)_{1j}\lambda_{j\alpha}\lambda^*_{1\alpha}\big]
\frac{1}{1-x_j}\bigg\}\,,
}
where $x_j\equiv M^2_j/M_1^2$ and
\eq{
g(x)\equiv \sqrt{x}\big[\frac{1}{1-x}+1-(1+x)\ln\big(\frac{1+x}{x}\big)\big]
\equiv \frac{\sqrt{x}}{1-x}+f(x)\,,
}
where $f(x)$ is the vertex function.
The part proportional to $f(x)$ corresponds to the one-loop vertex contribution
while the rest corresponds to the self-energy contribution for $N_R$.
We are assuming that $N_{iR}$ masses are hierarchical, i.e., $M_3-M_1>M_2-M_1\gg
\Gamma_1$. We comment on the possibility of resonant enhancement in
Sec.\,\ref{sec:resonant}.

Now if we apply the symmetry properties \eqref{prop:lambda} of $\lambda$ in
\eqref{eps:lambda}, we
conclude that
\eq{
  \label{cp:eps}
\eps_e=0\,,~~\eps_\mu=-\eps_\tau\,.
}
For example, note that $\lambda_{j\mu}^*=K^2_{Rjj}\lambda_{j\tau}$
and $K_{Rjj}^4=1$ for all $j$.
The $\cpmutau$ symmetry also relates the $\mu$ and $\tau$ washout parameters as
\eq{
  \label{cp:mtilde}
\tm_\mu=\tm_\tau\,,
}
where
\eq{
\tilde{m}_\alpha \equiv \frac{|\lambda_{1\alpha}|^2v^2}{M_1}
\,,
}
and $v=174\rm GeV$ in the SM;
they quantify the strength of $N_1$ decay and also its inverse decays into
$L_\alpha$.
Therefore, it is clear that the CP asymmetries for the $N_1$ decaying into all
flavors,
\eq{
  \label{eps.f}
\eps^{(1)}=\eps_e+\eps_\mu+\eps_\tau\,,
}
is vanishing and leptogenesis at the high scale $T\sim M_1\gtrsim 10^{12}\rm GeV$
can not proceed.

When $M_1\lesssim 10^{12}\rm GeV$, the tau Yukawa interactions enter in equilibrium
(also the muon flavor below $10^9$GeV) and distinct leptonic flavors may contribute
differently to leptogenesis.
In this case, the residual baryon asymmetry can be written
as\,\cite{flavored,nardi}
\eq{
  \label{YB:gen}
Y_{\Delta B}\simeq \frac{12}{37}\,Y^{\rm eq}_{N_1}\sum_\alpha
\eps_\alpha\eta_\alpha\,,
}
where the sum over $\alpha$ is performed only over the flavors that can be resolved
by interactions at the period of leptogenesis (one, two or three flavors).
The quantity $Y^{\rm eq}_{N_1}$ is the thermal density of $N_1$ per total entropy
density and is given by
$
Y^{\rm eq}_{N_1}=
\frac{135\zeta(3)}{4\pi^4g_*}\approx 3.9\times 10^{-3}\,,
$
where the last numerical value is for the SM degrees of freedom below the $N_1$
mass ($g_*=106.75$).
The factor $12/37$ corresponds to the reduction of asymmetry in
$\Delta_\alpha=\cB/3-\cL_\alpha$ to $\cB-\cL$ in the SM due to
spharelons\,\footnote{For the case of two Higgs doublets, this factor is $10/31$
but numerically very close.}.

When $10^9\lesssim M_1\lesssim 10^{12}\rm GeV$ only the $\tau$ Yukawa
interactions are in equilibrium and then only the $\tau$ flavor and its orthogonal
combination are resolved by interactions. In this case, the asymmetry in
\eqref{YB:gen} can be approximated by
\eq{
  \label{YB:2f}
Y_B\simeq \frac{12}{37}\times Y^{\rm eq}_{N_1}\times\bigg[
\eps_2\eta\Big(\frac{417}{589}\tilde{m}_2\Big)
+\eps_\tau \eta\Big(\frac{390}{589}\tilde{m}_\tau\Big)
\bigg]\,,
}
where $\eps_2=\eps_e+\eps_\mu$, $\tilde{m}_2=\tilde{m}_e+\tilde{m}_\mu$,
and
\eq{
\eta(\tilde{m}_\alpha)\simeq \bigg(
\Big(\frac{\tilde{m}_\alpha}{2.1m_*}\Big)^{-1}
+\Big(\frac{m_*/2}{\tilde{m}_\alpha}\Big)^{-1.16}
\bigg)^{-1}\,.
}
The mass $m_*\equiv \frac{16\pi^2v_u^2}{3\Mp }\sqrt{\frac{g_*\pi}{5}}
\approx 1\,\text{meV}$ quantifies the expansion rate of the Universe.
The factors 417/589 and 390/589 correspond to the diagonal entries of the $A$
matrix and quantifies the effects of flavor in the washout processes when changing
from the asymmetry in lepton doublets to asymmetries in $\Delta_\alpha$, see
e.g.\,\cite{nardi}.
We can see that the properties \eqref{cp:eps} of $\cpmutau$ leads to a partial
cancellation of the baryon asymmetry in \eqref{YB:2f} but it is nonzero because
the $\tau$ flavor and its orthogonal combination are washed out differently.
The question is then quantitative.
We show some cases leading to successful leptogenesis in
Sec.\,\ref{sec:lepto.results}.

For $M_1\lesssim 10^9\rm GeV$, the $\mu$ Yukawa interactions are also fast enough
so that the three flavors can be resolved.
For such a low scale, the CP asymmetries are usually too small to lead
to a successful leptogenesis.
 In the $\cpmutau$
symmetric case, the baryon asymmetry is in fact vanishing.
With the three flavors resolved, the baryon asymmetry can be approximated by
\eq{
  \label{YB:3f}
Y_B\simeq \frac{12}{37}\times Y^{\rm eq}_{N_1}\times\bigg[
\eps_e\eta\Big(\frac{151}{179}\tilde{m}_e\Big)
+\eps_\mu \eta\Big(\frac{344}{537}\tilde{m}_\mu\Big)
+\eps_\tau \eta\Big(\frac{344}{537}\tilde{m}_\tau\Big)
\bigg]\,.
}
Due to the properties \eqref{cp:eps} and \eqref{cp:mtilde}, the baryon asymmetry
vanishes within this analytic approximation.
Note that this is true even for mild hierarchies for $M_i$ and the leptogenesis
scale cannot be lowered by tuning the values of the masses.

Therefore, as long as $\cpmutau$ symmetry is valid at the leptogenesis scale, the
\textit{only} temperature range for which leptogenesis might be viable for
hierarchical $N_{iR}$ is the intermediate scale $T\sim M_1$ where
\eq{
10^{9}\text{GeV}\lesssim M_1\lesssim 10^{12}\text{GeV}\,.
}
It is worth emphasizing that  CP violation in our case comes from maximal Dirac CP
phase of the low-energy sector thereby giving a symmetry setting for some scenarios
of leptogenesis driven by low-scale CP violation\,\cite{pascoli}.
All these properties follow from the $G_l$ conservation in the charged lepton
sector and $\cpmutau$ conservation of the neutrino sector; see Sec.\,\ref{sec:sym}.

\subsection{Resonant leptogenesis}
\label{sec:resonant}

For the usual type-I seesaw scenario, the CP asymmetry produced by $N_1$ decay
usually decreases as we lower the mass of $N_1$ since the Yukawa couplings decrease
and {also the washout effects get stronger.}
For $M_1\ll 10^9\rm GeV$, successful leptogenesis is not possible for hierarchical
$N_{iR}$.
However, when some of the masses, say $M_1$ and $M_2$, are quasi-degenerate, it is
possible to resonantly enhance the CP asymmetry leading to the resonant
leptogenesis scenario\,\cite{resonant}.
In fact, \eqref{eps:lambda} is singular in that limit because
perturbation theory breaks down.
We can regulate such a behavior by resummation methods\,\cite{resonant}.
We will see in the following that $\cpmutau$ still leads to \eqref{cp:eps} and it
largely suppresses the CP asymmetries if $\mu$ and $\tau$ flavors have equal
washout strengths.

Suppose $M_3\gg M_2\approx M_1$ and also the resonant condition
\eq{
  \label{resonance}
M_2-M_1\sim \Gamma_{1,2}\ll M_{1,2}\,.
}
The resummed flavored CP asymmetry for $N_1\to L_\alpha+\phi$, neglecting $M_3$ and
vertex contributions, can be approximated by\,\cite{resonant} (see also \cite{dev})
\eq{
  \label{eps:resonant}
\eps^{(1)}_{\alpha}\approx
f_{\rm reg}^{12}\,
\frac{\im[(\lambda\lambda^\dag)_{21}\lambda_{1\alpha}^*\lambda_{2\alpha}]
+\frac{M_1}{M_2}
\im[(\lambda\lambda^\dag)_{12}\lambda_{1\alpha}^*\lambda_{2\alpha}]}
{(\lambda\lambda^\dag)_{11}(\lambda\lambda^\dag)_{11}}
\,,
}
where
\eq{
f_{\rm reg}^{12}\equiv \frac{(M_1^2-M_2^2)M_1\Gamma_2^{(0)}}
{(M_1^2-M_2^2)^2+(M_1\Gamma_2^{(0)})^2}\,.
}
One can see that \eqref{eps:resonant} is a regulated version of \eqref{eps:lambda},
neglecting the contribution of $f(x)$ (vertex) and regulating the function
$\sqrt{x_2}/(1-x_2)$ by $f_{\rm reg}^{12}$. See \cite{dev} for a discussion
about other regulator functions used in the literature.
The $N_2$ decay is also resonantly enhanced as
\eq{
\eps^{(2)}_\alpha\approx \eps^{(1)}_\alpha\,.
}
Thus with appropriate $\lambda$ we can have an enhanced CP asymmetry of order
one compared to $\eps\sim 10^{-6}$ required for successful leptogenesis in the
conventional case.

Now, since the Yukawa structure in \eqref{eps:resonant} is the same as in the
hierarchical case \eqref{eps:lambda}, the consequences of $\cpmutau$ are the same:
the flavored CP asymmetries $\eps^{(1)}_{\alpha},\eps^{(2)}_\alpha$ obey
\eqref{cp:eps}.
Therefore, if the effects of washout for $\mu$ and $\tau$ flavors are the same, the
CP asymmetries for $\mu$ and $\tau$ will cancel each other precluding leptogenesis
even when $M_1\sim M_2\lesssim 10^9\rm GeV$.
This would be the case in the analytic approximation \eqref{YB:3f} arising from the
classical Boltzmann equation solutions.
However, to properly quantify the baryon asymmetry, including washout effects, a
full flavored and quantum description is necessary and we will not address
it here.
Moreover, when the three right-handed neutrinos are quasi-degenerate, a more
complicated expression holds for the CP asymmetries\,\cite{dev} and it is not clear
if the properties \eqref{cp:eps} will still hold.

\subsection{Quantitative analysis and $N_3$ decoupled case}
\label{sec:lepto.results}

To assess quantitatively if leptogenesis can be successful with $G_F=G_l\times
G_\nu$ symmetry,
we can use the Casas-Ibarra parametrization that uses a complex orthogonal
matrix $R$:
\eq{
R=\hM_R^{-1/2}(\lambda v)U_\nu\hM_\nu^{-1/2}\,,
}
where the hatted matrices correspond to the diagonalized matrices and
$\lambda$ is in the basis \eqref{basis:lepto}.

We can see that the $\cpmutau$ symmetry implies
\eq{
R^*=K_R^2RK_\nu^2\,.
}
This means that there is no CP violating effect coming from $R$ when there is
$\cpmutau$ symmetry. A similar result was found for usual CP symmetry
in\,\cite{pascoli}.
CP invariance in $R$ is more apparent if we eliminate the potential purely
imaginary $i$ factors as in
\eq{
  \label{prop:R}
R=K_R^*\Rz K_\nu\,.
}
where $R^{(0)}$ is a real matrix, as can be seen from the properties of $R$.
Therefore, $\Rz$ obeys
\eq{
  \label{prop:R:2}
{\Rz}^\tp K_R^2\Rz=K_\nu^2\,,~~
{\Rz} K_\nu^2{\Rz}^\tp=K_R^2\,.
}
This is just the defining relation for a \textit{real} orthogonal matrix
when $K_R^2=K_\nu^2=\id$ or a \textit{real} hyperbolic\,\footnote{Lorentz
transformations in 2+1 dimensions.}
$\Rz$ in O(2,1), when $K_R^2=K_\nu^2=\diag(-1,1,1)$ or
any independently permuted diagonal entries for $K_R^2$ or $K_\nu^2$.
There is no other possibility and we conclude that the CP parities of
$\nu_{iL}$ ($N_{iR}$) are either all equal or only one is different.

When $M_i$ are hierarchical, the flavored CP asymmetries in \eqref{eps:lambda}
can be approximated to\,\cite{pascoli,nardi}
\eq{
  \label{eps.flavor:R}
\eps_{\alpha}= -\frac{3M_1}{16\pi v^2}
\frac{\im\{\sum_{ij} \sqrt{m_im_j}m_jR_{1i}R_{1j}U^*_{\alpha i}U_{\alpha j}
\}}
{\sum_j m_j|R_{1j}|^2}
\,,
}
where $M_1\ll M_2,M_3$ is assumed.
One can check \eqref{eps.f} also in this form from the properties for $R$ and
$U_{\alpha j}$ in Eqs.\,\eqref{prop:U} and \eqref{prop:R}.
Hence we only need $\eps_\tau$.

If we eliminate the CP parities $K_\nu,K_R$, we obtain
\eq{
  \label{eps.flavor:0}
\eps_{\tau}= -\frac{3M_1'}{16\pi v^2}
\frac{\sum_{ij}
\sqrt{m_im_j}m_j'\Rz_{1i}\Rz_{1j}\im\{{\Uz_{\tau i}}^*\Uz_{\tau j}\}}%
{\sum_j m_j(\Rz_{1j})^2}
\,,
}
where $M_1'=(K_R)^2_{11}M_1\equiv\pm M_1$ and $m_j'\equiv(K_\nu)^2_{jj}m_j=\pm m_j$
are the masses including the CP parities.
We can simplify further as
\eqali{
  \label{eps.flavor:1}
\eps_{\tau}&=
  \frac{3M_1'}{16\pi v^2\tm}\frac{J_\cp}{|U_{e1}U_{e2}U_{e3}|}
  \Big\{B_{12}\Rz_{11}\Rz_{12}-B_{13}\Rz_{11}\Rz_{13}
  +B_{23}\Rz_{12}\Rz_{13}
  \Big\}\,,
}
where
\eqali{
  \label{B.tma}
B_{ij}&\equiv \sqrt{m_im_j}(m_j'-m_i')|U_{ek}|\,,\cr
\tm&\equiv \sum_\alpha \tm_\alpha=\sum_j m_j|R_{1j}|^2=
\sum_\alpha\sum_{ij}\sqrt{m_im_j}\Rz_{1i}\Rz_{1j}
\re({\Uz_{\alpha i}}^*\Uz_{\alpha j})
\,,
}
with $(ijk)=(123)$ or permutations
and $J_\cp$ is the Jarlskog invariant
\eq{
J_{\cp}\equiv\im[U_{e1}U_{\mu2}U_{e2}^*U_{\mu1}^*]\,.
}
To obtain \eqref{eps.flavor:1}, we have multiplied and divided by
$\Uz_{11}\Uz_{12}\Uz_{13}=|U_{11}U_{12}U_{13}|$ and included the appropriate
factors inside the imaginary part. Notice that we are assuming $\cpmutau$  and
\eqref{U0}.
We also used the fact that the Jarlskog invariant can be written in terms of
different entries of $U$.

In the standard parametrization, the Jarlskog invariant is
\eq{
J_{\cp}=(s_{13}c_{13}^2)(s_{12}c_{12})(s_{23}c_{23})\sin\dcp\,.
}
Therefore, in the $\cpmutau$ symmetric case, we obtain
\eq{
  \frac{J_\cp}{|U_{e1}U_{e2}U_{e3}|}=\pm \frac{1}{2}\,,
}
for $\dcp=\pm \pi/2$, respectively\,\cite[b]{GL}.
We can see from \eqref{eps.flavor:1} that $\eps_\tau$ depends only on the
low-energy CP violation coming from $J_\cp$.
Other than that, $\eps_\tau$ only depends on the three $\Rz_{1i}$, on
the absolute neutrino scale and the discrete choice of $\nu_{iL}$ CP
parities.

We can finally use $Y_B$ in \eqref{YB:2f}, $\eps_\tau$ in \eqref{eps.flavor:1} and
$\tm_\alpha$ in \eqref{B.tma} to calculate the baryon asymmetry produced by
leptogenesis using the Casas-Ibarra parametrization.
To simplify the numerical study even further, we employ the approximation where
$M_{3}\gg M_{1,2}$ and $N_{3R}$ decouples.
In that case, the $R$ matrix can be written as\,\cite{N3:decoupling}
\eqali{
\text{NH:}&& R&=\mtrx{0&\star&\star\cr0&\star&\star\cr1&0&0},&&m_1\to0\,,
\cr
\text{IH:}&&R&=\mtrx{\star&\star&0\cr\star&\star&0\cr0&0&1},&&m_3\to0\,.
}
Then we can denote the different cases of CP parities for $N_{iR}$ and $\nu_{iL}$
as in Table\,\ref{tab.1}.
\begin{table}[h]
\eq{
\begin{array}{|c|c|c|c|}
\hline
 \text{Case} & K_R & K_\nu & \Rz \\
\hline
 (00) & \id_3 & \id_3 & \text{O(3)} \\
 (jk) & (K_R)_{jj}=i & (K_\nu)_{kk}=i & \text{O(2,1)} \\
\hline
\end{array}
}
\caption{\label{tab.1}
Possibilities for $K_R,K_\nu$ and $\Rz$. In cases $(jk)$, $j,k=1,2,3$,
$K_R,K_\nu$ have only one different diagonal entry as $\diag(i,1,1)$ or any
permuted diagonal entries.}
\end{table}
In the decoupling limit, when $R$ is not real, we only have the cases
[cf. \eqref{prop:R:2}]
\eqali{
\text{NH:}&& (31),(12),(13),(21),(23)\,;
\cr
\text{IH:}&& (33),(11),(12),(21),(22)\,.
}
Note that, differently from the strength of double beta decay,
leptogenesis also depends on the CP parities of the heavy right-handed neutrinos.

\begin{figure}[h]
\centering
\includegraphics[scale=0.37,angle=0]{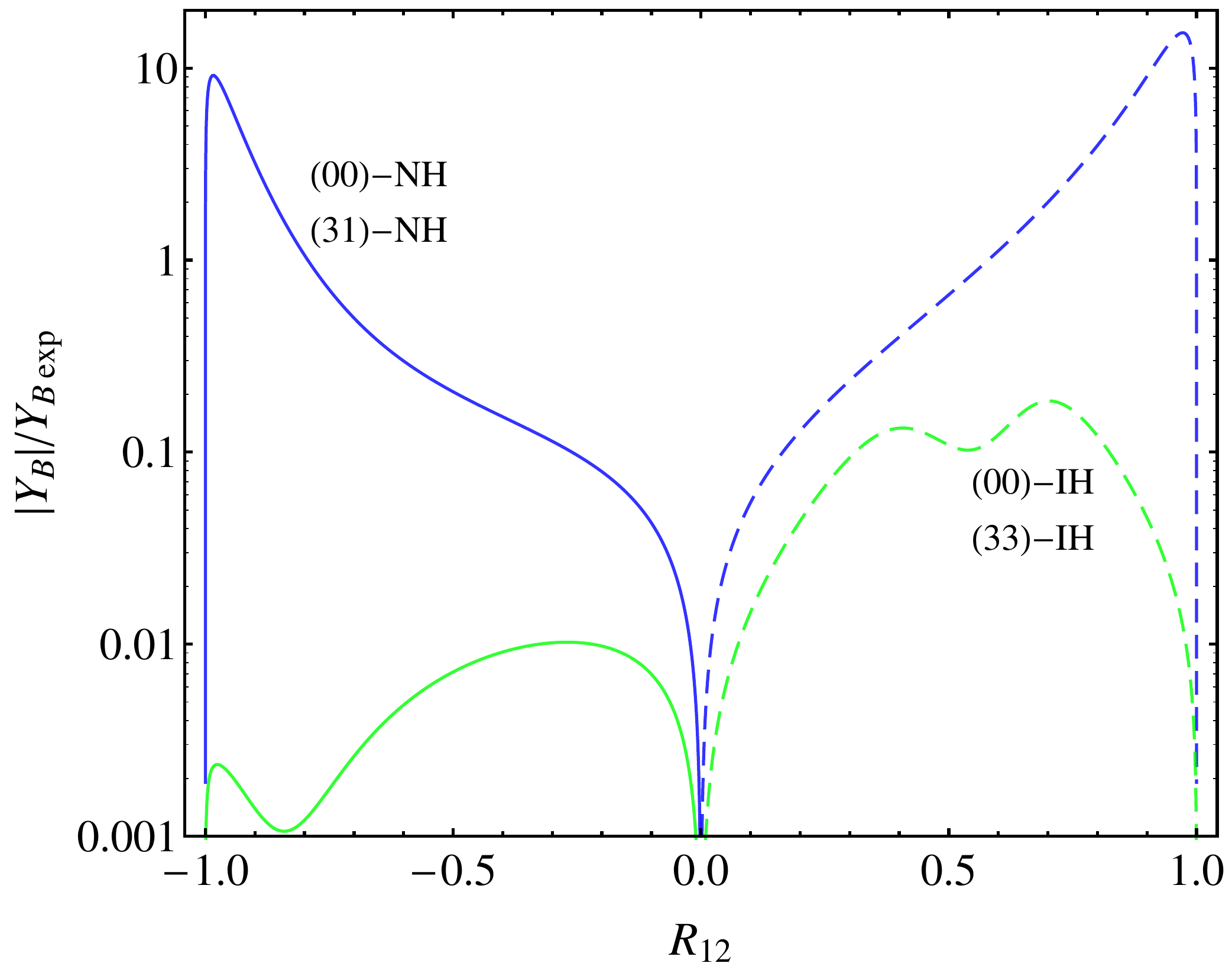}
\includegraphics[scale=0.37,angle=0]{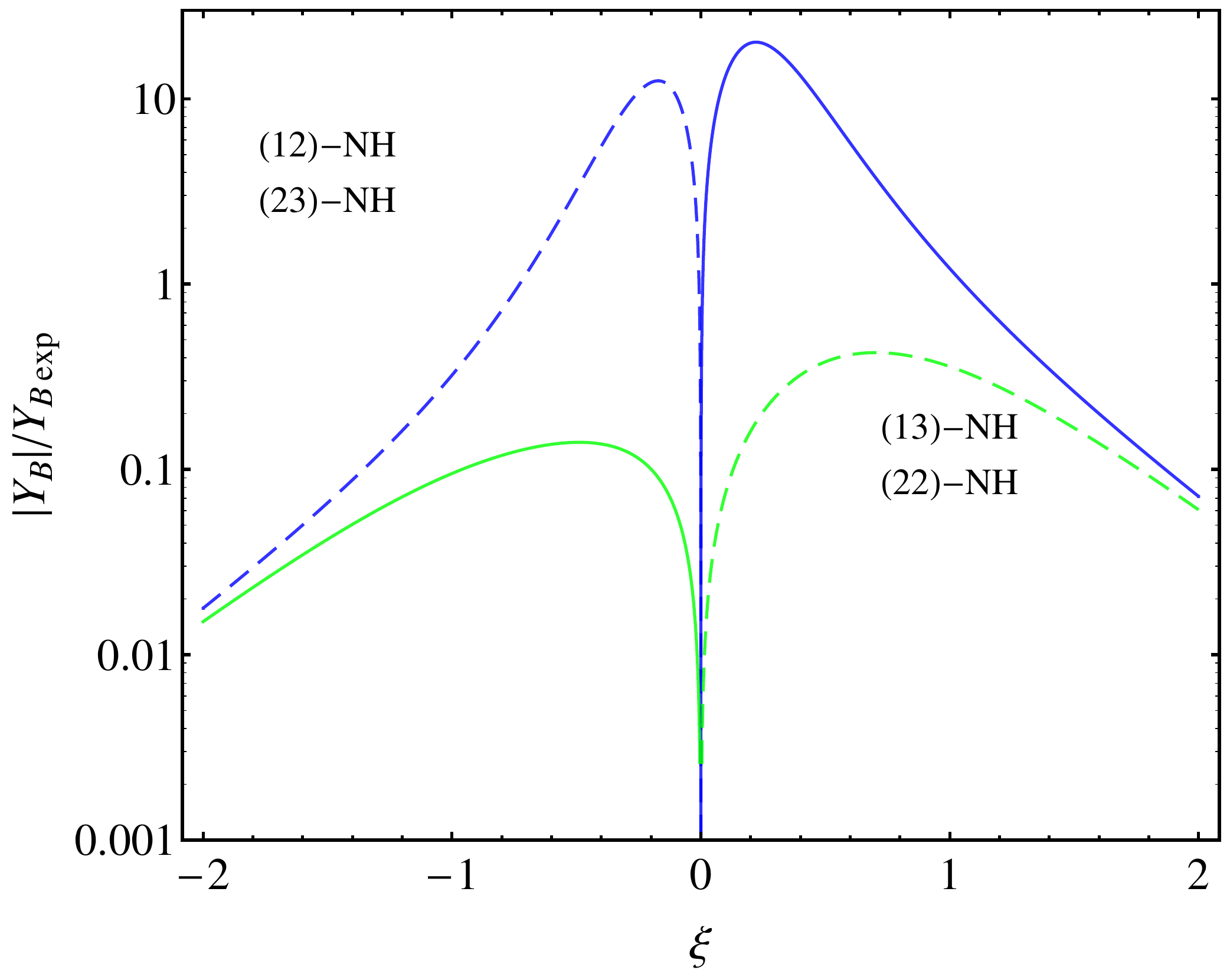}
\caption{\label{fig:lepto.1}
\textbf{Left}: ratio of $|Y_B|$ over $Y_{B\rm exp}=8.75\times 10^{-11}$ as a
function of $R_{12}$ for $M_1=10^{12}\rm GeV$ in the $N_{3R}$ decoupling limit;
the blue curve corresponds to both (00)-NH and (31)-NH, with
$R_{11}=0$, $|R_{12}|^2+|R_{13}|^2=1$ and $R_{13}>0$, while the green
curve corresponds to both (00)-IH and (33)-IH, with $R_{13}=0$,
$|R_{11}|^2+|R_{12}|^2=1$ and $R_{11}>0$.
\textbf{Right}: ratio of $|Y_B|$ over $Y_{B\rm exp}$,
for $M_1=10^{12}\rm GeV$ and in the $N_{3R}$ decoupled limit, as a function of
$\xi$
in $R_{1i}=(0,\cosh\xi,-i\sinh\xi)$ for (12)-NH (blue) and
$R_{1i}=(0,-i\sinh\xi,\cosh\xi)$ for (13)-NH (green);
the blue (green) curve also describes the case (23)-NH [(22)-NH], with
$R_{12},R_{13}$ exchanged and $\xi\to -\xi$.
We use the best-fit values of Ref.\,\cite{GG:fit} for $\theta_{12},\theta_{13}$ and
the squared mass differences.
The solid curves correspond to $Y_B>0$ for $\dcp=-90^\circ$ (preferred,
cf.\,\cite{valle:fit,GG:fit}) while the dashed curves correpond to $Y_B>0$ for
$\dcp=90^\circ$.
}
\end{figure}

\begin{figure}[h]
\centering
\includegraphics[scale=0.37,angle=0]{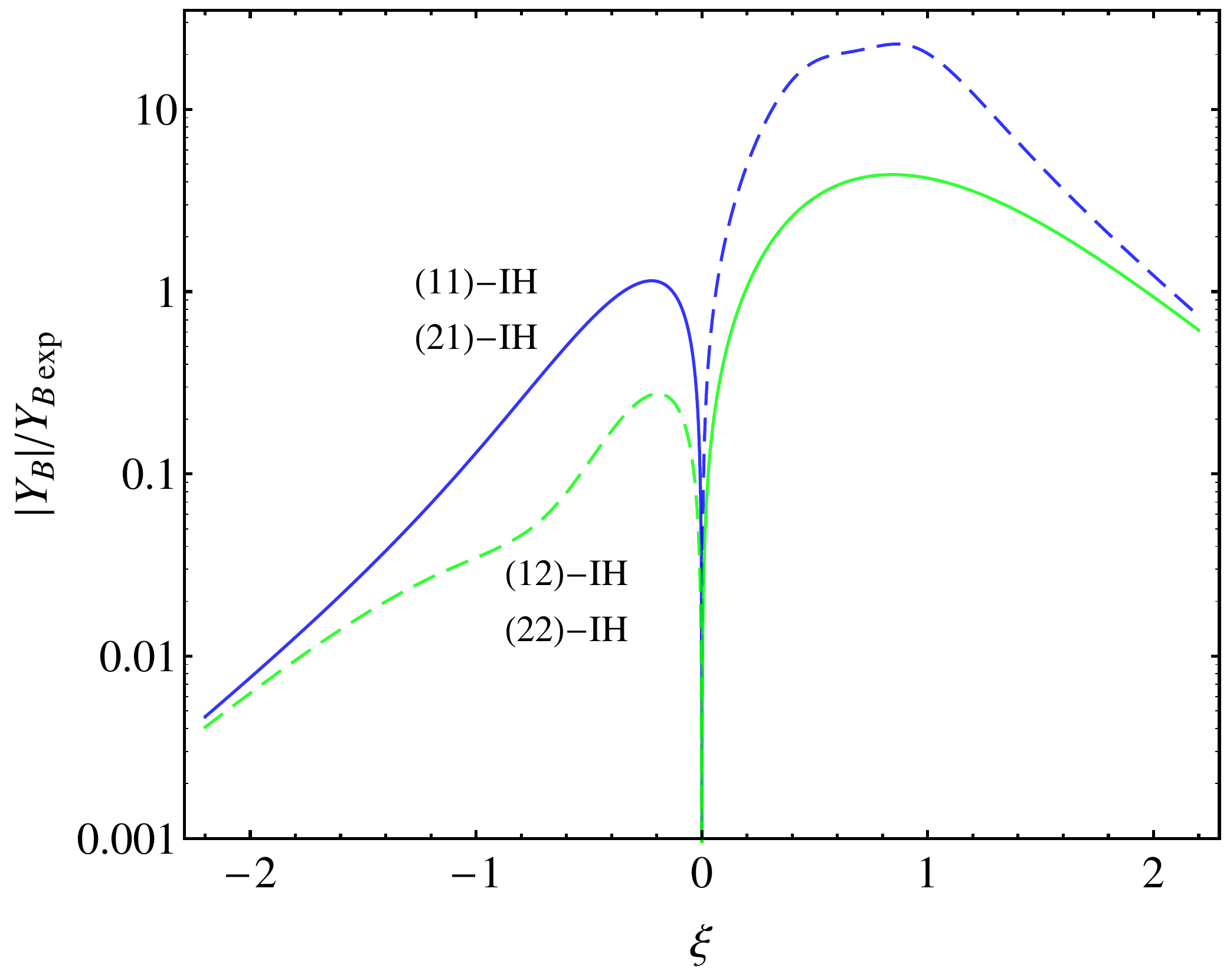}
\caption{\label{fig:lepto.2}
Ratio of $|Y_B|$ over $Y_{B\rm exp}=8.75\times 10^{-11}$ as a
function of $\xi$ for $M_1=10^{12}\rm GeV$ in the $N_{3R}$ decoupling limit;
$\xi$ is defined by $R_{1i}=(\cosh\xi,-i\sinh\xi,0)$ for (11)-IH (blue) and
$R_{1i}=(-i\sinh\xi,\cosh\xi,0)$ for (12)-IH (green);
the blue (green) also describes the case (21)-IH [(22)-IH], with
$R_{12},R_{13}$ exchanged and $\xi\to -\xi$.
The solid curves correspond to $Y_B>0$ for $\dcp=-90^\circ$ (preferred,
cf.\,\cite{valle:fit,GG:fit}) while the dashed curves correpond to $Y_B>0$ for
$\dcp=90^\circ$.
}
\end{figure}

We show our results for leptogenesis induced by hierarchical $N_{iR}$ and
decoupled $N_{3R}$ in Figs.\,\ref{fig:lepto.1} and \ref{fig:lepto.2}.
We use the maximum possible value for $M_1$ within flavored leptogenesis with
$\tau$-flavor in equilibrium: $M_1=10^{12}\rm GeV$. Given the parametrization in
\eqref{eps.flavor:R} ($M_1$ only appears linearly in the prefactor), lowering $M_1$
leads to proportional lowering of $\eps_\tau$ and also $|Y_B|$.
Plots with smaller $M_1$ can be obtained by scaling down the lines proportionally.
Note that $\theta_{23}=45^\circ$ (and $\dcp=\pm\pi/2$) is fixed from symmetry and
this makes the curves of $|Y_B|$ smoother, with less possibility of cancellations.

Let us begin with Fig.\,\ref{fig:lepto.1}, left.
We treat the case where all CP parities are equal for light and heavy
neutrinos, i.e., cases (00)-NH and (00)-IH, and the figure shows the ratio of the
baryon asymmetry of the model over its experimental value, $Y_B/Y_{B\rm exp}$, in
terms of $R_{12}$. Since the third $N_{3R}$ decouples, the same plots also applies
to the case where the CP parity of $N_{3R}$ is different from the rest, i.e.,
$K_R=\diag(1,1,i)$. The property in \eqref{prop:R:2} requires that we are only left
with the cases (31)-NH [same as (00)-NH] and (33)-IH [same as (00)-IH].
Thus successful leptogenesis can happen for normal hierarchy [(00)-NH and (31)-NH]
but not for the inverted hierarchy [(00)-IH and (33)-IH].
For normal hierarchy, we can read from the plot that the scale of $M_1$ can be
lowered at most by a factor $|Y_B|_{\max}/Y_{B\rm exp}= 15.3$ and we need
$0.65\times 10^{11}\lesssim M_1\lesssim 10^{12}\rm GeV$.
A similar situation of leptogenesis induced solely by $\dcp$ was also considered in
Ref.\,\cite{pascoli}. Here we furnish a symmetry justification for that case.

In Fig.\,\ref{fig:lepto.1}, right, the remaining cases for NH are considered, i.e.,
(12)/(23) and (13)/(22). We show the ratio $|Y_B|/Y_{B\rm exp}$ in terms of $\xi$,
which parametrizes the nonzero $R_{1i}$.
The cases (12) and (23) [(13) and (22)] are
represented by the same blue (green) curve. We can see that
the cases (13)-NH and (22)-NH do not lead to successful leptogenesis.
For (12)-NH and (23)-NH, successful leptogenesis is also possible for
$0.5\times10^{11}\lesssim M_1\lesssim 10^{12}\rm GeV$
($|Y_B|_{\max}/Y_{B\rm exp}=20.2$).

Finally, Fig.\,\ref{fig:lepto.2} shows the remaining cases for IH: (11)/(12) and
(12)/(22). We show again the ratio $|Y_B|/Y_{B\rm exp}$ in terms of $\xi$,
which parametrizes the nonzero $R_{1i}$.
In all cases leptogenesis is possible for slightly different ranges for $M_1$.
For (11)/(21), we need $0.44\times10^{11}\lesssim M_1\lesssim 10^{12}\rm GeV$
($|Y_B|_{\max}/Y_{B\rm exp}=22.8$).
For (12)/(22), $2.3\times10^{11}\lesssim M_1\lesssim 10^{12}\rm GeV$
($|Y_B|_{\max}/Y_{B\rm exp}=4.4$).
If we assume negative $\dcp$, preferred from global fits\,\cite{valle:fit,GG:fit},
then the range for case (11)/(21) shrinks almost to the single value $M_1\approx
10^{12}\rm GeV$ because the right portion of the curve leads to anti-matter
dominance instead of matter dominance; see figure.

We conclude that successful leptogenesis is not possible for the cases (00)-IH,
(33)-IH, (13)-NH and (22)-NH. Therefore, for IH, successful leptogenesis requires
that the CP parity of $\nu_{1L}$ or $\nu_{2L}$ be different of the rest.
On the other hand, the cases (00)-IH and (33)-IH correspond to the largest value of
$|m_{ee}|$ in Fig.\,\ref{fig:pred:mee}. If this value of $|m_{ee}|$ were measured
in future experiments, then $\cpmutau$ symmetric leptogenesis with hierarchical
right-handed neutrinos and decoupled $N_{3R}$ is excluded as the origin of the
present baryon asymmetry of the Universe.

%%%%%%%%%%%%%%%%%%%%%%%%%%%%%%%%%%%%%%%%%%%%%%%%%%%%%%
\section{Symmetry choice and properties}
\label{sec:sym}

We now turn to a theoretical discussion of $\cpmutau$ symmetry and follow it up in the subsequent section with a model realization. As already noted, a much pursued idea in the neutrino literature is that flavor symmetries may be behind the structure of masses
and mixing angles of the leptons~\cite{review}.
A very predictive setting consists of assuming that the charged lepton sector and
neutrino sectors are invariant under different groups $G_l$ and $G_\nu$,
respectively. These groups are then part of a larger group $G_F$ that may be
 entirely or partially valid at higher energies (the latter if some factor appears
accidentally).
A less ambitious variations of the above idea is (i) to allow
more free parameters by requiring less symmetry for $G_\nu$ or $G_l$ or (ii)
including generalized CP (GCP) symmetries as part of the flavor group.
Here we pursue a direction where we identify a minimal setting with $G_l$ being
abelian and $G_\nu$ being a GCP transformation. We find that we are largely
restricted to $\cpmutau$ for $G_\nu$.

To discuss our strategy, we assume Majorana neutrinos, with the leptonic Lagrangian below EWSB in the flavor basis
to be
\eq{
\label{lag:mass}
-\lag=m_\alpha\bar{l}_{\alpha L}l_{\alpha R}
+\overline{\nu^c_{\alpha L}}(M_{\nu})_{\alpha \beta}\nu_{\beta L}
+h.c.\,,
}
where the implicit sum over $\alpha=e,\mu,\tau$  is understood.
Note that in the flavor basis, the interaction with $W$ gauge bosons is diagonal,
$W_\mu\bar{l}_{\alpha L}\gamma^\mu \nu_{\alpha L}$, and the PMNS matrix comes from the
diagonalization of $M_\nu$.

It is clear that the charged lepton part of \eqref{lag:mass} is invariant under
three separate family lepton numbers $\cL_e,\cL_\mu,\cL_\tau$, that
should be broken in the neutrino part.
Although these symmetries are automatically present whenever we diagonalize the
charged lepton mass matrix\,\cite{grimus.ludl}, we assume some subgroup of it,
$G_l$, is a symmetry of the theory at higher scales for the charged lepton sector
(we allow for the fact that it may be accidental). Since charged leptons and
left-handed neutrinos come from the same leptonic doublet $L_\alpha$ above the EW
scale, the group $G_\nu$ should also act on the same space.
Let us look for the minimal $G_l$ and $G_\nu$ where the former is abelian and the
latter is a GCP.

We assume $G_l$ has a generic element acting on $L_\alpha=(l_{\alpha L},\nu_{\alpha
L})$ of the form (more generic forms are considered in appendix \ref{ap:proof})
\eq{
  \label{T}
G_l:\quad T=\mtrx{1&&\cr&e^{i\theta}&\cr&&e^{-i\theta}}\,.
}
For the moment, $G_l$ can be a continuous $U(1)$ group
{(which can therefore be the group $\Umutau$ of $\cL_\mu-\cL_\tau$)} or
a discrete abelian group $\ZZ_n$, with $n\ge 3$ to avoid degenerate $T$.
We are in the basis where $T_{L}=T_{l_R}=T$ act all in the same way on left-handed
doublets and right-handed singlets but they can be in different irreducible
representations (irreps) if $T$ is embedded in a larger group. In this case, $G_l$
will refer to the group acting on the left-handed doublets $L_\alpha$.

Next, we assume the symmetry of the neutrino sector of \eqref{lag:mass}, $G_\nu$,
is composed of a generalized CP (GCP) symmetry\,\cite{gcp} of the form
\eq{
  \label{gcp:X}
G_\nu:\quad L(x)\to XL^{\cp}(\hx)\,,
}
where $L^{\cp}=-iCL^*$ is the usual CP transformation and $X$ is a generic
$3\times 3$ unitary and \textit{symmetric} matrix acting in the space of three
families; $\hx$ is the space inversion of $x$. Symmetric $X$ guarantee that the
application of \eqref{gcp:X} two times, leads to the identity.
Note that a global rephasing is unimportant for $X$.

Now we \textit{demand} that $G_l$ and $G_\nu$ \textit{close} as a group acting on
$L_\alpha$. If $G_{l,\nu}$ were unitary and we demanded that the product of its
generators
be finite, we would obtain von Dyck groups that were extensively studied in
this context\,\cite{hernandez}.%
\footnote{
For a different approach based on $\ZZ_2\times\ZZ_2$, see \cite{z2z2}.
}
Instead, \eqref{gcp:X} is a GCP symmetry and we should demand that the
following composition of $G_\nu$ and $G_l$ induce an
automorphism\,\cite[d]{fcp}:
\eq{
  \label{cond.1}
XT^*X^\dag=T'\in G_l\,.
}
where $T, T'$ are elements of the same group. This equation can be rewritten as
\eq{
    \label{cond.2}
X=T'XT^\tp\in G_l\,.
}
This equation and the previous one are not restricted to diagonal $T$ but
are valid for any unitary $T$ in any basis.

If $G_l=U(1)$,
{irrespective of the form in Eq.\eqref{T}}, there are only
two possible autormorphisms:
\eq{
  \label{auto}
(i)~~T'=T^{-1} \quad \text{ or }\quad
(ii)~~T'=T \,.
}
These are also automorphisms for all subgroups $\ZZ_n$ and, in particular,
for $n=3,4$, they are the only ones.
For general $\ZZ_n$, with $n\neq 3,4$, additional automorphisms $T'=T^k$ are
possible but not with the form \eqref{T}.
For these automorphisms, \eqref{cond.2} and the form of $T$ in \eqref{T} leads to
\eq{
  \label{X}
(i)~~X=\mtrx{1 &&\cr&1&\cr&&1} \quad \text{ or }\quad
(ii)~~X=\mtrx{1 &0&0\cr0&0&1\cr0&1&0}\,,
}
after rephasing some fields appropriately.
The first case is just usual CP transformation and we can see that the charged lepton
part of \eqref{lag:mass} is automatically invariant under such a transformation,
thus
leading to CP invariance in the whole theory.
This symmetry prevents CP violation in the leptonic sector and hence we consider it
no further. Instead we focus on the
second case which we will denote as $\cpmutau$ and it is a well-known GCP
symmetry in the literature called $\mu\tau$-reflection
symmetry\,\cite{GL}.
CP breaking arises in this setting because of the clash between the neutrino
part and the charged lepton part in \eqref{lag:mass}:
the former is invariant under $\cpmutau$ while the latter is invariant under the
usual CP (after rephasing).
What distinguishes our work from the previous ones is that in previous works on $\cpmutau$, neither the symmetry $G_l$ was
identified nor its relation with $G_\nu$ was stressed as we do here.
Also, in later approaches using GCP symmetry with finite flavor
symmetries, much more complicated automorphism structures (compared to ours)  needed
to be studied for some groups\,\cite{fcp,fcp:others}.

In fact, this settings is much more general: the two forms for $X$ in \eqref{X} are
\textit{unique} for any diagonal $T$ and the form for $T$ in \eqref{T} is also
\textit{unique} for $G_l=U(1)$ or $G_l=\ZZ_n$ with prime $n$ or $n=4,6$.
The uniqueness is up to simultaneous permutations of rows and columns that leaves
$T$ diagonal.
This result is proved in appendix \ref{ap:proof}, where we also show the first
different form for $T$ -- it occurs for $\ZZ_8$.

Permutations of the above structure can be discarded for phenomenological reasons
as follows.
If we adopt $T$ with nontrivial entries in (11)-(22) [or (11)-(33)], the
structure of $X$ would also be interchanged and we obtain the relations
$|U_{e3}|=|U_{\mu 3}|$ (or $|U_{e3}|=|U_{\tau 3}|$), which leads (respectively) to
\eqali{
\cp^{e\mu}:&& \tan\theta_{13}=\sin\theta_{23}\,,\cr
\cp^{e\tau}:&& \tan\theta_{13}=\cos\theta_{23}\,.
}
These relations are completely excluded because of small $\theta_{13}$.

At last, we point out a remarkable property of the symmetries $G_l$ generated by
$T$ and $G_\nu=\zcp$ generated by $\cpmutau$: the two groups
commute.\,\footnote{This property is more transparent in the basis where $G_l$, in
the continous case, is represented by $SO(2)$ rather than $U(1)$ and $\cpmutau$ is
represented by usual CP which commutes;
see appendix \ref{ap:real}.
}
Therefore, our minimal flavor group, including GCP, can be just $G_F=G_l\times
G_\nu$.\,\footnote{Obviously $\cpmutau$ may not commute with other symmetries
such as the SM gauge group.}
Generically, when $G_F$ is a subgroup of U(3), $G_l\sim
\ZZ_n$ and $G_\nu\sim \ZZ_2\times\ZZ_2$ (or subgroup), their commutation is
impossible because all mixing angles are nonzero.
For that reason, the whole group containing $G_l$ and $G_\nu$ tends to be a large
nonabelian group. For example, the minimal group that leads to TBM is
$S_4$\,\cite{lam:s4} of order 24. To fix at least the nonzero $\theta_{13}$, it
must be much larger of order 150 or more\,\cite{group-scans}.

The commutation of $G_l$ and $G_\nu$ seems to have another remarkable feature,
i.e., the \textit{vev alignment problem}\,\footnote{
This name is not entirely appropriate in our context (we use one-dimensional
irreps, see also appendix\,\ref{ap:real}) and we specifically refer here to the
possibility of different symmetry breaking scalars interacting through the
potential.
}
often encountered in flavor symmetry model building --
can be naturally avoided in the scalar sector (without supersymmetry)
as our examples below show.
The solution is simply that $G_l$ ($G_\nu$) can be broken in the neutrino sector
(charged lepton sector) preserving $G_\nu$ ($G_l$) by using $G_\nu$-invariant
($G_l$-invariant) fields with $G_l$ ($G_\nu$) charge.
Hence, only complete invariants of both $G_l$ and $G_\nu$ interact in the potential.
Thus to avoid the contamination of $G_l$-breaking effects in the neutrino sector,
we just need to avoid the coupling of $G_l$ breaking scalars to neutrino fields
(be it by additional symmetries).
The same is valid for the charged lepton sector.\footnote{To see the advantage of
our discussion relative to other flavor groups, we
can compare our setting with those based on $A_4=(\ZZ_2\times\ZZ_2)\rtimes\ZZ_3$
group. We can take $G_l\simeq \ZZ_3$ and $G_\nu\simeq \ZZ_2$ and note that they do
not commute.
In this case, $G_\nu$ invariant fields with $G_l$ charge exist: take the
$\bs{1}'$ or $\bs{1}''$
singlets (in actual models, additional flavons are necessary to partly break
$\ZZ_2\times\ZZ_2$ of $A_4$).
However, there is no irrep with $G_\nu$ charge
but without $G_l$ charge in $A_4$. In actual $A_4$ models, usually triplets
$\bs{3}$ with specifically aligned vevs are used to achieve the breaking $G_F\to
G_l$ in the charged lepton sector (and also in the neutrino sector, hence the
alignment problem).
}

%%%%%%%%%%%%%%%%%%%%%%%%%%%%%%%%%%%%%%%%%%%
\section{Model}
\label{sec:model}

The main challenge in model building with $\cpmutau$, is to keep it
unbroken in the neutrino sector while breaking it sufficiently in the charged
lepton sector (keeping $G_l$) to generate $\mu$-$\tau$ mass splitting.
We have found several ways to meet this challenge.
Although our general setting can be implemented in many different ways, some distinction
is possible on how $G_l$ appears and how $G_\nu$ (GCP) is broken in the charged
lepton sector.
The different possibilities depend on how $G_l$  appears, i.e., either
\begin{itemize}
\item $G_l$ comes from a symmetry of the whole theory $G_F$ at high scales; or
\item $G_l$ appears accidentally.
\end{itemize}
In Sec.\,\ref{sec:sym}, we saw that the largest group for abelian $G_l$ is
$G_l=\Umutau$, which is the continuous symmetry of the combination
$\cL_\mu-\cL_\tau$.
Variations on this respect involve gauging $\Umutau$ or considering only a $\ZZ_n$
subgroup of it. The latter would allow embedding our $G_l\times G_\nu$ into
a larger nonabelian discrete group.
Either way, we use the nomenclature of $\Umutau$ to describe our models and only
make some comments on variants.

Furthermore, our setting requires that only $G_l$ be broken in the neutrino sector
and only $G_\nu$ be broken in the charged lepton sector -- the conservation of $G_l$
and $G_\nu$ in the complementary sectors is what leads to predictions. That is
achieved through the vacuum expectation value of scalars that we call as
$l$-flavons and $\nu$-flavons. They have the following properties:
\begin{itemize}
\item $l$-flavons: all conserve $G_l$ but some need to break $G_\nu$. Best
candidate is a $G_l$ invariant $\cpmutau$ \textit{odd} scalar (we denote it as
$\sigma$).
\item $\nu$-flavons: all conserve $G_\nu$ but some need to break $G_l$.
Best candidates are scalars carrying $G_l$ charge but $\cpmutau$
\textit{even} ($G_\nu$-invariant); we denote them as $\eta$'s.
\end{itemize}
Since the alignment problem in the scalar potential can be avoided, we just need to
prevent $l$-flavons ($\nu$-flavons) to couple to the neutrino sector (charged
lepton sector).
Often that can be achieved by additional symmetries.

One remark with respect to additional symmetries of flavons is in order.
For the above setting, it is simpler if flavons do not carry other additive quantum
numbers other than those of $G_l$ or $G_\nu$.
For example, let us consider a $\nu$-flavon $\eta_2$ carrying
$\cL_\mu-\cL_\tau=2$ ($G_l$) so that it couples with $N_3$ as $N_{3R}^2\eta_2$.
If $\eta_2$ carries no other quantum number, we can
define its $\cpmutau$ transformation
as\,\footnote{
This possibility is raised for general discrete nonabelian symmetries in
\cite[d]{fcp} but no model application was discussed.
}
\eq{
  \label{cp:eta2}
\cpmutau:\quad
\eta_2(x)\to \eta_2(\hx)\,,
}
i.e., $\eta_2$ is composed of two CP \textit{even} real scalars.
However, if $\eta_2$ also carries $\cB-\cL=-2$, and $N_{iR}$ carries $\cB-\cL=-1$,
then its $N_3$ coupling transform as
\eq{
\label{N3.eta2}
  \cpmutau:\quad
N_{3R}N_{3R}\eta_2\to N_{2R}^{cp}N_{2R}^{cp}\eta_2\,,
}
which maps a $\cB-\cL$ invariant term to a $\cB-\cL$ violating term.
In this case, consistency with $\cpmutau$ requires the existence of another field
$\eta_{-2}$ with charges $\cL_\mu-\cL_\tau=-2$ and $\cB-\cL=-2$. The transformation
property now would be
\eq{
\label{cp:eta2-2}
  \cpmutau:\quad \eta_2(x)\to \eta_{-2}^*(\hx)\,.
}
This corrects the transformation properties for \eqref{N3.eta2} but allow
$\cpmutau$ breaking if $|\aver{\eta_2}|\neq |\aver{\eta_{-2}}|$.
Therefore, in our setting, we require that $\nu$-flavons carry no other additive
quantum number and hence a continuous $\cB-\cL$ symmetry cannot be implemented.

The exception to the above feature is when $\nu$-flavons carry
only a $\ZZ_2$ quantum number. In this case, since the representation is real,
\eqref{cp:eta2} can be maintained.
This means that a $\ZZ_4$ subgroup of $U(1)_{\cB-\cL}$, acting as
\eq{
  \label{Z4L}
\ZZ^L_{4}:\quad
\text{leptons}\sim i\,,\quad
\text{$\nu$-flavons}\sim -1\,,
}
can still be implemented as a symmetry.

At last, we assume leptogenesis is successful in our setting and we will be seeking
high scale ($M_R\gtrsim 10^{11}\rm GeV$) type-I seesaw  implementations.

\subsection{Multi-Higgs implementation}
\label{sec:mHiggs}

The model below illustrates the general aspects of our setting.
In this case, $G_l$ will be accidental and $G_\nu$ will be broken at a high scale
and transmitted to the charged lepton sector to generate the $\mu\tau$ mass
splitting.
The symmetries at the high scale will be $\Umutau\times\zcp$, a gauged
$\Umutau$ (which is not exactly $G_l$ at the low scale) and global $\cpmutau$.
Another implementation where $G_F=G_l\times G_\nu$ is a symmetry of the high
scale theory is given in appendix \ref{sec:heavy.lep}.

All lepton fields transform alike under $\Umutau$, with $\cL_\mu-\cL_\tau$ charges
\eq{
  \label{Umutau:Li}
L_i\sim l_i\sim N_i\sim (0,1,-1)\,,
}
where $L_i,l_i\equiv l_{iR},N_i\equiv N_{iR}$ (here we use $L_i,l_i$ instead of
$L_\alpha,l_\alpha$) are the three families of lepton doublets, lepton singlets and
right-handed neutrino singlets, respectively.
The $\cpmutau$ symmetry also acts similarly for all of the
three type of fields, as \eqref{gcp:X} with the second $X$ in \eqref{X},
and should swap the second with the third family fields.
Note that this GCP symmetry commutes with $\Umutau$ and it does not reverse its
charges. The SM group charges, however, are reversed by this GCP symmetry.

We add two more Higgs doublets $\phi_{\pm 2}$ with $\Umutau$ charge $\pm 2$ in
addition to the SM doublet $\phi_0$.
The Lagrangian for the charged lepton sector is
\eq{
  \label{lag:l}
-\lag^l=y_0\bar{L}_1\phi_0 l_1+y_2\bar{L}_2\phi_2l_3+y_{-2}\bar{L}_3\phi_{-2}l_2\,.
}
We prevent $\phi_0$ from coupling to $\bar{L}_2l_2$ and $\bar{L}_3l_3$ by assigning
\eq{
\ZZ_2:\quad l_2,l_3,\phi_{\pm 2} \text{ are odd}.
}
Such a symmetry also leads to the accidental symmetry
\eq{
G_l:\quad
L_2\sim l_3\sim e^{i\theta}\,,\quad
L_3\sim l_2\sim e^{-i\theta}\,.
}
The Higgs doublets are invariant under this symmetry and so it leaves the
symmetry invariant even after EWSB. It is this symmetry that will correspond to
$G_l$ at low energies and will differ from our original $\Umutau$ only for $l_{iR}$.
The $\cpmutau$ acts in the same form for $G_l$ as it does for $\Umutau$.

The $\cpmutau$ symmetry acts on the doublets as
\eq{
\phi_0\to \phi_0^*\,,\quad
\phi_2\to \phi_{-2}^*\,.
}
This implies $y_0$ is real and $y_{-2}^*=y_2$.

If we write
\eq{
\aver{\phi_{-2}^{(0)}}=v_{-2}~~\text{and}~~\aver{\phi_2^{(0)}}=v_2\,,
}
the $\cpmutau$ breaking will come from
\eq{
  \label{v22}
|v_{-2}|\gg |v_2|\,,
}
which induces the $\mu\tau$ mass splitting
\eq{
m_\mu=|y_2 v_2|~\ll~ m_\tau=|y_{-2}v_{-2}|\,.
}
Note that prior to EWSB $\cpmutau$ renders $\mu\tau$ flavors indistinguishable and
the $|v_{-2}|\ll |v_2|$ leads physically to the same situation.
The $\cpmutau$ breaking in \eqref{v22} will be induced by a large vev of a CP odd
scalar $\sigma$ in the potential~\cite{CMP}.

The Higgs potential is
\eqali{
V_2&= \mu_2(|\phi_2|^2+|\phi_{-2}|^2)+\mu_0|\phi_0|^2
\,,\cr
V_4&=
\ums{2}\lambda_0|\phi_0|^4+\ums{2}\lambda_1(|\phi_2|^2+|\phi_{-2}|^2)^2
+\lambda_2|\phi_2|^2|\phi_{-2}|^2
\cr&~~
+\lambda_{22}(\phi_0^\dag\phi_2\phi_0^\dag\phi_{-2} +h.c.)
+\lambda_{02}|\phi_0|^2(|\phi_2|^2+|\phi_{-2}|^2)
\cr&~~
+\lambda_{02}'(|\phi_0^\dag\phi_2|^2+|\phi_0^\dag\phi_{-2}|^2)
\,,
\cr
\delta V&=\mu_\sigma\sigma(|\phi_2|^2-|\phi_{-2}|^2)
+(\lambda_{-4}\phi_2^\dag\phi_{-2}\eta_2^2+h.c.)
}
where $\sigma$ is a CP odd scalar and $\eta_2$ is a CP-even scalar with $\Umutau$
charge 2 and will couple to $N_2^2,N_3^2$.
We have omitted a term similar to the $\lambda_2$-term because only neutral vevs are
sought and they are not relevant to the discussion below.
We could also replace $\Umutau$ by $\ZZ_8$ by adding the terms
$(\phi_2^\dag\phi_{-2})^2$.

After $\sigma$ and $\eta_2$ acquire vevs at the high scale, we get from $\delta
V$ and $V_2$ an effective quadratic term for $\phi_{\pm 2}$,
\eq{
V_{2\rm eff}=
M_2^2|\phi_2|^2+M_{-2}^2|\phi_{-2}|^2+M^2_{22}\phi_2^\dag\phi_{-2}+h.c.,
}
where
\eq{
M_2^2=\mu_2+\mu_\sigma\aver{\sigma}\,,\quad
M_{-2}^2=\mu_2-\mu_\sigma\aver{\sigma}\,,\quad
M_{22}^2=\lambda_{-4}\aver{\eta_2}^2\,.
}
Irrespective of the phases of $\lambda_{-4},\aver{\eta_2}$, we apply rephasing
transformations so that $M_{22}^2$ is real and negative.

Now we adjust $\aver{\sigma}$ so that $|M_2^2|\simeq \veps^{-1} |M_{22}^2|\simeq
\veps^{-2} |M_{-2}^2|\sim
v_{\rm ew}$. The phases of the vevs are trivial in the minimum when $\lambda_{22}<0$.
This leads to a high scale mass matrix for $(\phi_{-2}, \phi_{2}) $ of the form:
\begin{eqnarray}
M^2_\phi~=~M^2_2\left(\begin{array}{cc}\veps^2  & \sim\veps \\ \sim\veps & 1\end{array}\right)
\end{eqnarray}
The two approximate eigenvectors of this matrix are: $H^\prime\approx \phi_2+\veps
\phi_{-2}$ and $h_0\approx \phi_{-2}-\veps \phi_{2}$. By fine tuning we keep
$\veps\sim \frac{m_\mu}{m_\tau}$ and $H^\prime$ as superheavy whereas $h_0$ mass is
negative and weak scale. Then below the scale of $\aver{\eta}$ and $\aver{\sigma}$,
the effective charged lepton Yukawa couplings in \eqref{lag:l} look like:
\eq{
  \label{eff:yuk:2hdm}
-\lag^l_{\rm eff}\simeq y_0\bar{L}_1\phi_0 l_1+y_2\veps \bar{L}_2 h_0 l_3
+y^*_2\bar{L}_3 h_0l_2\,.
}
After a $90^\circ$ rotation of the right-handed charged leptons, this gives $m_\tau
=|y_2^*\aver{h_0^{(0)}}|$ and $m_\mu=|y_2\veps \aver{h_0^{(0)}}|$ as desired for a
realistic theory.

For the neutrino sector we add three singlet scalars $\eta_k$, $k=0,1,2$,
with $\Umutau$ charge $k$; $\eta_0$ is a real scalar.
When they acquire vevs (for $k\neq 0$), they break $\Umutau$ without breaking
$\cpmutau$, as discussed previously, and they transform trivially under $\cpmutau$:
\eq{
\cpmutau:\quad \eta_k(x)\to \eta_k(\hx)\,,
}
where $\hx=(x_0,-\bx)$ for $x=(x_0,\bx)$.
We also assume the symmetry $\ZZ_4^L$ in \eqref{Z4L} where
$\eta_k\sim -1$.

The Lagrangian for $N$,
\eqali{
  \label{lag:NN}
-\lag&\supset \ums{2}k_1\bar{N}_1N_1^c\eta_0+k_{23}\bar{N}_2N_3^c\eta_0\cr
&\quad +\
\ums{2}k_2\bar{N}_2N_2^c\eta_2+\ums{2}k_3\bar{N}_3N_3^c\eta_2^*
\cr
&\quad +\
k_{12}\bar{N}_1N_2^c\eta_1+k_{13}\bar{N}_1N_3^c\eta_1^*
\,,
}
gives rise to $M_R$ in the $\cpmutau$ symmetric form \eqref{mutau-r:form} after
$\eta_k$ acquire generic vevs.
GCP symmetry imposes real $k_1$, real $k_{23}$, $k_3=k_{2}^*$, $k_{13}=k_{12}^*$.
Given the necessary structure \eqref{mutau-r:form} and the requirement for
$\theta_{13}\neq 0$, we indeed need both fields $\eta_{1,2}$.
Note that $\ZZ_4^L$ prevents $\sigma$ from coupling to $N_{iR}$.

It can be seen that $\cpmutau$ symmetric $M_R$ also leads to a $\cpmutau$ symmetric
$M_R^{-1}$.
Such a structure is maintained from the neutrino Dirac mass matrix $M_D$ coming from
\eq{
\label{lag:LN}
-\lag\supset f_0\bar{N}_1\tphi_0^\dag L_1+
f_2\bar{N}_2\tphi_0^\dag L_2+f_3\bar{N}_2\tphi_0^\dag L_3\,,
}
where $\phi_0$ is the same Higgs doublet that couples to electrons and quarks.
The reality of $f_0$ and $f_3=f_2^*$ follow from $\cpmutau$ and and we
obtain
\eq{
M_D=\mtrx{x_\nu&&\cr &z_\nu&\cr &&z_\nu^*}\,.
}
The neutrino mass matrix given by the seesaw formula~\cite{seesaw}
\eq{
  \label{seesaw}
M_\nu=-M_D^\tp M_R^{-1}M_D\,,
}
is $\cpmutau$ invariant and has the form \eqref{mutau-r:form} as
advertised.

The leptogenesis aspects studied in Sec.\,\ref{sec:lepto} has to be adapted in
this case because $v=174\rm GeV$ has to be replaced by $v_u=v\sin\beta$.
The plots shown in Figs.\,\ref{fig:lepto.1} and \ref{fig:lepto.2} apply now
for $M_1/\sin^2\beta=10^{12}\rm GeV$ and limits for the $M_1$ window changes
accordingly.

\subsection{Higgs spectrum}

At low energies, the scalar sector of this model acts like a
lepton-specific (also called type-X) two Higgs doublet
model\,\cite{branco} with the Higgs doublets being $h_0$ and $\phi_0$,
except for the Higgs couplings to electrons; cf.\,\eqref{eff:yuk:2hdm}.
Both of the doublets acquire vevs such that
$\sqrt{\aver{\phi_0}^2+\aver{h_0}^2}=v=174$ GeV. The ratio of vevs is given by
$\aver{h_0}/\aver{\phi_0}=\tan\beta$ and the mixing between the real neutral Higgs
fields is denoted by $\tan \alpha$. The effective Higgs potential in terms of
$\phi_0$ and $h_0$ is given by:
\eqali{
V(\phi_0,h_0)&= -\mu^2_\phi |\phi^2_0|-\mu^2_h|h_0|^2
+\ums{2}\lambda_0|\phi^2_0|^2+\ums{2}\lambda_1|h^2_0|^2
\cr&~~
+\lambda_{02}|\phi_0|^2|h_0|^2
+\lambda'_{02}|\phi_0^\dag h_0|^2
+\lambda_{22}\veps(\phi_0^\dag h_0\phi_0^\dag h_0 + h.c.)\,.
}

The spectrum of Higgs states is given by\,\cite{2hdm:m}
\eq{
m^2_{A}=-4\lambda_{22}\veps v^2\,,\quad
m^2_{H^+}=-(\lambda_{02}'+2\veps\lambda_{22})v^2\,,
}
where $v=174\,\rm GeV$ (we use a different normalization
compared to \cite{2hdm:m}), while the mass matrix for the CP even states,
in the basis $\sqrt{2}(\re h_0-v_{-2},\re\phi_0-v_0)$, is
\eq{
M^2_{h,H}=
2\mtrx{\lambda_1v_{-2}^2&\lambda_{345}v_0v_{-2}\cr
      \lambda_{345}v_0v_{-2}&\lambda_0v_0^2}
\,,
}
where $\lambda_{345}=\lambda_{02}+\lambda_{02}'+2\veps\lambda_{22}$.
We are using $\aver{h_0}\approx \aver{\phi_{-2}}\approx v_{-2}$.

Since our parameter $\lambda_{22}$ comes from the high energy theory (decoupled
$\phi_2$), it can not be arbitrarily large.
If we impose it to be perturbative,
$|\lambda_{22}|<4\pi$ we obtain an upper bound for the pseudoscalar $A$ as
\eq{
m_A=2v\sqrt{\veps|\lambda_{22}|}\lesssim
2v\sqrt{4\pi\frac{m_\mu}{m_\tau}}
\approx 300\,\text{GeV}\,,
}
hence non-decoupling.
This is smaller than $2m_t$ and $t\bar{t}$ cannot be produced.
Neutral scalars in the 2HDMs are less constrained than the charged
higgsses (e.g. from flavor observables\,\cite{2hdm:flavor}) and the strongest limits
are available for the MSSM (or type-II)\,\cite{pdg}.
Usually they appear as lower bounds on the heavy masses because the decoupling
limit is usually a good description.
Very light pseudoscalars of mass below $\mathcal{O}(10\rm GeV)$ can also have
its couplings constrained\,\cite{pdg,light.A}.
Current LHC limits for the different types of 2HDM constrain the
various 2HDMs to be close to the alignment limit\,\cite{craig}.
Even in this limit, a portion of the parameter space is already excluded. For
example, only $\tan\beta\gtrsim 3$ is allowed by data (above $200\rm
GeV$).
Also, being an effective 2HDM, the triple Higgs coupling for the interaction
$h^3$ is different from the SM and can be probed in the future\,\cite{nir}.

\section{Summary}
\label{sec:summary}

We have presented a minimal setting where $G_l$ is conserved in the charged
lepton sector and $G_\nu$ is conserved in the neutrino sector.
The largest $G_l$ can be identified with the combination $\cL_\mu-\cL_\tau$ symmetry
and $G_\nu$ is generated by a generalized CP symmetry, $\cpmutau$, that combines CP
with $\mu\tau$ exchange.
When $G_l$ is conserved in the charged lepton sector and $G_\nu$ is conserved in
the neutrino sector, we obtain the usual prediction of maximal $\theta_{23}$ and
$\dcp$ with nonzero $\theta_{13}$.
Additionally, Majorana phases are fixed up to discrete choices and they
lead to very specific predictions for neutrino-less double beta decay and
leptogenesis.

In our setting, the two symmetries $G_l$ and $G_\nu$ commute and this feature
allows us to naturally avoid the alignment problem in the scalar sector.
Additional symmetries can be used to keep the $G_l$- and $G_\nu$-breaking effects
restricted to the neutrino sector and charged lepton sector, respectively.
Additionally, continuous $\cB-\cL$ cannot be imposed (hence not gauged) in our
setting
and only a $\ZZ_4$ subgroup may be imposed to keep $\cpmutau$ naturally
unbroken in the neutrino sector.
Our construction also illustrates that generalized CP symmetries based on the
\textit{trivial} automorphism of flavor groups -- much less considered in the
literature -- may still lead to interesting model constructions.

For the neutrino-less double beta decay, the discrete choice of Majorana phases
(or CP parities) leads to specific strips that can be clearly distinguished in some
cases; see Fig.\,\ref{fig:pred:mee}.
For example, for inverted hierarchy, the case of all equal CP parities or
only $\nu_{3L}$ with different CP parity can be distinguished from the rest and
can be potentially measured or falsified in the near future. We emphasize that, key predictions of these models are: (i) $\theta_{23}=45^0$ and $\delta_{CP}=\pi/2$
simultaneously i.e.  if experimentally measured
values for either of these observables deviate from the above predictions, $\cpmutau$ violating terms will be necessary to keep these ideas viable.

The consequences of $\cpmutau$ for leptogenesis leads to the natural implementation
of the purely flavored leptogenesis scenario where the total CP asymmetry due to
$N_1$ decay is vanishing.
Successful leptogenesis is possible only when flavored leptogenesis is
considered and that must take place at the intermediate temperature range of
$10^{9}$--$10^{12}\rm GeV$. Flavored leptogenesis below $10^9\rm GeV$ seems to be
precluded even if the CP asymmetry is resonantly enhanced by quasi-degenerate
$N_{1R}$
and $N_{2R}$ if the $\mu$- and $\tau$-flavors are washed out equally.
For effective two heavy and hierarchical right-handed neutrinos the window for
successful leptogenesis is even narrower: $5\times 10^{10}$--$10^{12}\rm GeV$.

\section*{Acknowledgements} The work of R.N.M. is supported
in part by the National Science Foundation Grant No. PHY-1315155.
C.C.N is partially supported by Brazilian Fapesp grant 2013/26371-5 and
2013/22079-8. C.C.N. is also grateful for the hospitality of the Maryland Center
for Fundamental Physics.

%%%%%%%%%%%%%%%%%%%%%%%%%%%%%%%%%%%%%%%%%%%

%%%%%%%%%%%%%%%%%%%%%%%%%%%%%%%%%%%%%%%%%%%
\appendix
\section{Uniqueness of $\cpmutau$}
\label{ap:proof}

We show here that the GCP defined by $X$ in \eqref{X} for the abelian symmetry
generated by $T$ in \eqref{T} are the only possibilities for any $G_l=U(1)$ or
$G_l=\ZZ_n$, with prime $n$ or $n=4,6$. A different possibility arises only for $T$
(the possibilities for $X$ are the same) beginning with $n=8$.
The case of $G_l=U(1)$ was considered in the text. We only need to consider
$G_l=\ZZ_n$.

To show the assertion, we generalize the form of $T$ from \eqref{T} to
\eq{
T= \mtrx{z_1&&\cr &z_2&\cr&&z_3}\,,
}
where $z_i$ are complex number of modulus unity. We also keep $\det T=z_1z_2z_3=1$
because its nontrivial contribution can be factored out to usual lepton number.
Let us also consider more general automorphisms for $\ZZ_n$ in \eqref{cond.2}:
$T'=T^k$ where $k$ cannot divide $n$.

Then the consistency condition \eqref{cond.2} can be recast in the following form:
\eq{
  \label{cond.3}
z_i^kz_j=1 ~~\text{ if }~~ X_{ij}\neq 0\,.
}
Let us take the first row of $X$. Because $X$ is nonsingular, at least one element
of the first row has to be nonzero. Suppose two elements are nonzero. If
$X_{11}\neq 0$ and $X_{12}\neq 0$, then condition \eqref{cond.3} implies
\eq{
z_1^{k+1}=z_1^kz_2=1\,,
}
and then $z_1=z_2$ which is impossible because $T$ is nondegenerate. The same
conclusion is reached if any two of the elements of the first row is nonzero.
The argument is independent of the row and hence only one element in each row (or
column) can be nonzero.
Listing all possibilities and selecting only the symmetric matrices, the nonzero
entries of $X$ coincides with the positions of the unity in \eqref{X}, after
eliminating similar forms that are related by the simultaneous permutations of rows
and columns that keep $T$ diagonal. Rephasing of fields leads to \eqref{X}.
Thus $X$ is restricted to \eqref{X} except for permutations.

Now, for the first case of $X=\id$, we reach the conclusion that
\eq{
z_1^{k+1}=z_2^{k+1}=z_3^{k+1}=1\,.
}
This means $T^{k+1}=\id$ and if $T$ is a faithful representation, $k+1=0\mod n$.
Therefore, $k=-1$ is the only possibility.

For the second case of $X$ being (23)-transposition, we have the conditions
\eq{
z_1^{k+1}=z_2^kz_3=z_3^kz_2=1\,.
}
This imposes conditions on $z_1$ and also $z_2^{k-1}=z_3^{k-1}$.
For prime $n$ the last relation is only possible if $k=1$: this leads to
\eqref{T} (we exclude $z_2=z_3$).
The cases $n=4,6$ do not lead to different forms because only $k=1$ or
$k=-1$ correspond to automorphisms.
The cases so far proves the assertion.

The first different form appears for $\ZZ_8$ and one example is
\eq{
T=\mtrx{-1&&\cr &\om_8&\cr &&\om_8^3}\,,
}
which obeys $XT^*X^{-1}=T^5$; $\om_8$ denotes $e^{i2\pi/8}$.
If we allow $X$ to be nonsymmetric, other possibilities appear such as for $\ZZ_7$
where $X$ is the cyclic permutation  and $T=\diag(\om_7,\om^2_7,\om^4_7)$ (the same
that appears for the $T_7$ group).

%%%%%%%%%%%%%%%%%%%%%%%%%%%%%%%%%%%%%%%%%%%
\section{$G_l\times G_\nu$ in the real basis}
\label{ap:real}

We show here how the $\cpmutau$ symmetry of \eqref{gcp:X} and the $\Umutau$
symmetry of \eqref{T} are rewritten in a real basis where $\cpmutau$ is just the
usual CP transformation. In this basis, the commutation of $\cpmutau$ and $\Umutau$
is transparent and it also shows how the combination $\Umutau\times \zcp$
leads effectively to a two-dimensional representation when the fields are
complex, i.e., carrying quantum numbers other than $\Umutau\times\zcp$.

It is clear that the charged lepton part of \eqref{lag:mass} breaks the $\cpmutau$
symmetry strongly as $m_\tau/m_\mu\sim y_\tau/y_\mu\sim 17$ (if $l_\alpha$ transform
similarly to $L_\alpha$ and $H$ transforms as usual).
This breaking can be analyzed in a different basis.
Since the matrix $X$ in $\cpmutau$ is symmetric, there is a change of basis
where $X$ can be completely removed. We can concentrate in the $\mu\tau$ space
where such a basis change is
\eq{
\mtrx{L_\mu\cr L_\tau}=\frac{1}{\sqrt{2}}\mtrx{1&-i\cr 1&i}
\mtrx{L_\mu'\cr L_\tau'}\,.
}
For the right-handed singlets $l_i$ we apply the same transformations.
The CP transformation in the new basis will be just the usual
\eq{
  \label{cp:usual}
L_i'\to (-iC)L_i'^*\,,
}
and similarly for $l_i$.

The Yukawa coefficients in $\bar{L}_iY_{ij}H l_j$ in the new basis will be just
\eq{
  \label{Y:so2}
Y=\mtrx{y_e&&\cr &y_\mu&\cr &&y_\tau}\to ~
Y'=\mtrx{y_e&&\cr &\bar{y}& -i\Delta y/2\cr&i\Delta y/2&\bar{y}}\,,
}
where $\bar{y}\equiv (y_\tau+y_\mu)/2$ and $\Delta y\equiv y_\tau-y_\mu$.
One can see that if $y_\tau\neq y_\mu^*$, CP is violated because the phases of
$\bar{y}$ and $i\Delta y$ can not be simultaneously removed [keeping \eqref{so2}]
while in this basis $M_\nu$ should be a real matrix.
For example, if $y_{\mu,\tau}$ are real CP is violated by $i\Delta y$.
The latter term is however still invariant by the following $SO(2)$ without being
proportional to the identity:
\eq{
  \label{so2}
\mtrx{L_\mu'\cr L_\tau'}\to
\mtrx{\cos\theta & \sin\theta\cr
-\sin\theta & \cos\theta }
\mtrx{L_\mu'\cr L_\tau'}\,,
}
In this basis it is clear that the CP transformation \eqref{cp:usual} commutes with
the $SO(2)$ transformation in \eqref{so2}.

In this basis it is also clear $\Umutau\times\zcp$ have the irreducible
representations shown in table \ref{tab:irreps},
\begin{table}
\eq{
\begin{array}{|c|c|c|}
\hline
\text{irrep} & \text{real basis} & \text{$\Umutau$ diagonal} \cr
\hline
(0,\pm ) &\text{1-dim real}& \text{1-dim real}\cr
(q,\pm ) &\text{2-dim real}& \text{1-dim complex}\cr
(0,*) &\text{1-dim complex}& \text{1-dim complex}\cr
(q,*) &\text{2-dim complex}& \text{2-dim complex}\cr
\hline
\end{array}
}
\caption{\label{tab:irreps}
Irreducible representations for $\Umutau\times\zcp$.}
\end{table}
where $(q,\pm)$ denotes charge $q$ for $\Umutau$ and CP parities $\pm$ while $*$
denotes a complex field transforming as $\phi\to \phi^*$ in the real basis or
$\phi_q\to \phi_{-q}^*$ in the $\Umutau$ diagonal basis.

%%%%%%%%%%%%%%%%%%%%%%%%%%%%%%%%%%%%%%%%%%%
\section{Single Higgs implementation}
\label{sec:heavy.lep}

In this implementation, the symmetry at the high scale is $G_F=G_l\times G_\nu$
where $G_l=\Umutau$ (gauged) and $G_\nu=\zcp$. At low energy, right
above the electroweak scale, we effectively have the SM with one Higgs doublet.

The neutrino sector is the same as in the multi-Higgs model of
Sec.\,\ref{sec:mHiggs}, with additional simplification by eliminating
$\eta_0$ and the symmetry $\ZZ_4^L$.
If we replace $\Umutau$ by $\ZZ_3$, we can simplify further by identifying
$\eta_1=\eta_2^*$, and we are left with only one $\nu$-flavon.

The charged lepton sector needs to be modified. We still assume $\cpmutau$ is
spontaneously broken by a vev of a CP odd scalar, which now we rename as $\sigma_-$.
We also need a CP even scalar $\sigma_+$.
To confine the CP breaking to the charged lepton sector, we
introduce a $\ZZ_2$ symmetry for which
\eq{
  \label{Z2}
\ZZ_2:\quad \sigma_{\pm}, l_{iR}~\text{ are odd},
}
and the rest are even. Both $\sigma_{\pm}$ are invariant under $\Umutau$.
We can write an effective Lagrangian as
\eq{
  \label{lag:l:eff}
-\lag^l_{\rm eff}=\frac{\sigma_e}{\Lambda_{\cp}}\bar{L}_eHl_e
+\frac{\sigma_\mu}{\Lambda_{\cp}}\bar{L}_\mu Hl_\mu
+\frac{\sigma_\tau}{\Lambda_{\cp}}\bar{L}_\tau Hl_\tau
+h.c.
\,.
}
where $\sigma_\alpha$, $\alpha=e,\mu,\tau$ are some complex linear combinations of
$\sigma_{\pm}$.
GCP invariance requires
\eqali{
  \label{sigma:a}
\sigma_e&=a_e\sigma_+ +ib_e\sigma_-\,,\cr
\sigma_\mu&=a_\mu\sigma_+ +ib_\mu\sigma_-\,,\cr
\sigma_\tau&=a_\tau\sigma_+ +ib_\tau\sigma_-\,,
}
where $a_e,b_e$ are real coefficients and $a_\tau=a_\mu^*$, $b_\tau=b_\mu^*$ are
generally complex.
The $\mu\tau$ mass splitting is generated from
\eq{
\frac{m^2_\tau-m^2_\mu}{v^2}=
\frac{1}{\Lambda_{\cp}^2}\Big[|a_\mu^*u_+ +ib_\mu^* u_-|^2-|a_\mu u_+ +ib_\mu u_-|^2
\Big]
  =\frac{u_+u_-}{\Lambda_{\cp}^2}4\im(a_\mu^*b_\mu)\,,
}
where $\aver{\sigma_+}=u_+$ and $\aver{\sigma_-}=u_-$. We can see that CP breaking,
and hence $\mu\tau$ mass splitting, requires both $u_{\pm}$ to be nonzero.

One example for a UV completion of \eqref{lag:l:eff} can be achieved by introducing
three heavy vector-like charged lepton fields $E_{iL}$ and $E_{iR}$, the latter with
the same SM quantum number of $l_{iR}$. They are charged under $\Umutau$ just like
the rest of the leptons as \eqref{Umutau:Li} but they are even under the additional
$\ZZ_2$ symmetry of \eqref{Z2}.
The Lagrangian is then
\eqarr{
  \label{lag:l}
-\lag^l&=y_1'\bar{L}_1HE_{1R}+y_2'\bar{L}_2HE_{2R}+y_3'\bar{L}_3HE_{3R}
\cr&\ +
M_{E_i}\bar{E}_{iL}E_{iR}+\sigma_i\bar{E}_{iL}l_{i}\,,
}
where $y_3'=y_2'^*$ and $\sigma_i$ are some linear combinations of $\sigma_{\pm}$
just like \eqref{sigma:a}; $M_{E_1}$ is real from GCP and $M_{E_3}=M_{E_2}$ can
be taken real by convention.
We obtain \eqref{lag:l:eff} for the charged leptons after integrating out the heavy
leptons $E_{i}$, with the identification
\eqali{
  \label{sigma.alpha}
\frac{\sigma_e}{\Lambda_\cp}&=-\frac{y_1'}{M_{E_1}}\sigma_1\,,\cr
\frac{\sigma_\mu}{\Lambda_\cp}&=-\frac{y_2'}{M_{E_2}}\sigma_2\,,\cr
\frac{\sigma_\tau}{\Lambda_\cp}&=-\frac{y_3'}{M_{E_3}}\sigma_3\,.
}
In particular, the electron Yukawa is naturally small for
$M_{E_1}\gg M_{E_2}$.

We should mention that $\Umutau$ breaking would be induced in the charged lepton
sector by the additional couplings between $E_i$ and $\eta_k$ as
\eqali{
-\lag^l&\subset
  \mu_{12}'\bar{E}_{1L}E_{2R}\eta_1^*
  +\mu_{13}'\bar{E}_{1L}E_{3R}\eta_1
\cr&\quad
  +\mu_{21}'\bar{E}_{2L}E_{1R}\eta_1
  +\mu_{31}'\bar{E}_{1L}E_{3R}\eta_1^*
\cr&\quad
  +\mu_{23}'\bar{E}_{2L}E_{3R}\eta_2
  +\mu_{32}'\bar{E}_{3L}E_{2R}\eta_2^*
+h.c.,
}
where $\mu'_{32}=\mu'^*_{23}$, $\mu'_{13}=\mu'^*_{12}$, $\mu'_{31}=\mu'^*_{21}$.
However, we can assume that $\Umutau$ breaking scale is much smaller than the bare
mass terms for $E_i$ as
\eq{
  \label{scales.1}
|\aver{\eta_{1,2}}|\ll M_{E_2}\ll M_{E_1}\,.
}
In this case, the $\Umutau$ breaking effects can be neglected and \eqref{lag:l:eff}
is effectively obtained after $E_i$ are integrated out.
Since $\aver{\eta_k}$ are related to $N_R$ masses, more specifically to the
generation of $\theta_{12},\theta_{13}$ and $N_2,N_3$ mass splitting,
\eqref{scales.1} means that $N_R$ mass scale is much smaller than the $E_i$
scale.
An alternative way of avoiding $\Umutau$ breaking in the charged lepton
sector would be to use $\ZZ_4^L$.

As for the scale of $\aver{\sigma_{\pm}}$, we should have
$\aver{\sigma_{\pm}}/M_{E_2}\gtrsim 10^{-2}$ for an order one $y_3'$
coupling in \eqref{sigma.alpha},
and it can lie below or above the $\Umutau$
breaking scale. Anyhow, $\sigma_{\pm}$ does not couple to $N_R$ at renormalizable
level due to the $\ZZ_2$ symmetry and CP breaking is not induced at leading order
to the neutrino sector since $\eta_{k}$ only couple to CP even
combinations $\sigma_+^2$ and $\sigma_-^2$.
We assume, however, that all $\aver{\eta_k}$, $\aver{\sigma_{\pm}}$, are greater
than the scale where leptogenesis takes place, typically $10^{11}\rm GeV$ in our
case, so that CP breaking in the charged lepton sector can be manifest.

%%%%%%%%%%%%%%%%%%%%%%%%%%%%%%%%%%%%%%%%%%%

%%%%%%%%%%%%%%%%%%%%%%%%%%%%%%%%%%%%%%%%%%%%%%%%%

\section*{\Large References}
\begin{thebibliography}{99}
% \setlength{\baselineskip}{.98\baselineskip}

\bibitem{valle:fit}
  D.~V.~Forero, M.~Tortola and J.~W.~F.~Valle,
%   ``Neutrino oscillations refitted,''
  Phys.\ Rev.\ D {\bf 90} (2014) 9,  093006
  [\xlink{1405.7540}];

\bibitem{GG:fit}
  M.~C.~Gonzalez-Garcia, M.~Maltoni and T.~Schwetz,
%   ``Updated fit to three neutrino mixing: status of leptonic CP violation,''
  JHEP {\bf 1411} (2014) 052
  [\xlink{1409.5439}].

\bibitem{lepto_rev}  M.~Fukugita and T.~Yanagida,
  %``Baryogenesis Without Grand Unification,''
  Phys.\ Lett.\ B {\bf 174}, 45 (1986);  W.~Buchmuller, P.~Di Bari and M.~Plumacher,
  %``Leptogenesis for pedestrians,''
  Annals Phys.\  {\bf 315}, 305 (2005)
  [\xlink{hep-ph/0401240}];

\bibitem{pascoli} S.~Pascoli, S.~T.~Petcov and A.~Riotto,
  %``Connecting low energy leptonic CP-violation to leptogenesis,''
  Phys.\ Rev.\ D {\bf 75}, 083511 (2007)
  [\xlink{hep-ph/0609125}];
%   S.~Pascoli, S.~T.~Petcov and A.~Riotto,
%   ``Leptogenesis and Low Energy CP Violation in Neutrino Physics,''
  Nucl.\ Phys.\ B {\bf 774} (2007) 1
  [\xlink{hep-ph/0611338}];
  G.~C.~Branco, R.~Gonzalez Felipe and F.~R.~Joaquim,
  %``A New bridge between leptonic CP violation and leptogenesis,''
  Phys.\ Lett.\ B {\bf 645} (2007) 432
  [\xlink{hep-ph/0609297}].
% \bibitem{lepto:2}
For a further discussion on this issue, see
  S.~Davidson, J.~Garayoa, F.~Palorini and N.~Rius,
  %``Insensitivity of flavoured leptogenesis to low energy CP violation,''
  Phys.\ Rev.\ Lett.\  {\bf 99} (2007) 161801
  [\xlink{0705.1503}].



\bibitem{review} See for example, G.~Altarelli and F.~Feruglio,
  %``Discrete Flavor Symmetries and Models of Neutrino Mixing,''
  Rev.\ Mod.\ Phys.\  {\bf 82}, 2701 (2010)
  [\xlink{1002.0211}];
S.~F.~King and C.~Luhn,
  %``Neutrino Mass and Mixing with Discrete Symmetry,''
  Rept.\ Prog.\ Phys.\  {\bf 76}, 056201 (2013)
  [\xlink{1301.1340}].


\bibitem{mutauexc}
T.~Fukuyama and H.~Nishiura,
  %``Mass matrix of Majorana neutrinos,''
  \xlink{hep-ph/9702253};
R.~N.~Mohapatra and S.~Nussinov,  Phys.\ Rev.\ {\bf D 60}, 013002 (1999)
  [\xlink{hep-ph/9809415}];
E.~Ma and M.~Raidal,
  %``Neutrino mass, muon anomalous magnetic moment, and lepton flavor
  % nonconservation,''
  Phys.\ Rev.\ Lett.\ {\bf 87}, 011802 (2001); erratum ibid. 87,
  159901 (2001) [\xlink{hep-ph/0102255}];
C.S. Lam,
%A 2Ð3 symmetry in neutrino oscillations,
  Phys. Lett. {\bf B 507}, 214 (2001) [\xlink{hep-ph/0104116}];
K.~R.~S.~Balaji, W.~Grimus, and T.~Schwetz,
%The solar LMA neutrino oscillation solution in the Zee model,
Phys.\ Lett.\ {\bf B 508}, 301 (2001)
[\xlink{hep-ph/0104035}];
  E. Ma, %The all-purpose neutrino mass matrix,
  Phys. Rev. D 66, 117301 (2002) [\xlink{hep-ph/0207352}];
  A. Ghosal, %A neutrino mass model with reflection symmetry,
  Mod. Phys. Lett. A 19, 2579 (2004);
  R.~N.~Mohapatra,
 % ``theta(13) as a probe of mu <---> tau symmetry for leptons,''
  JHEP {\bf 0410} (2004) 027 [\xlink{hep-ph/0408187}].
  T.~Kitabayashi and M.~Yasue,
  %``mu-tau symmetry and maximal CP violation,''
  Phys.\ Lett.\ B {\bf 621}, 133 (2005) [\xlink{hep-ph/0504212}];
R.~N.~Mohapatra and W.~Rodejohann,
  %``Broken mu-tau symmetry and leptonic CP violation,''
  Phys.\ Rev.\ D {\bf 72}, 053001 (2005) [\xlink{hep-ph/0507312}];
E.~I.~Lashin, N.~Chamoun, C.~Hamzaoui and S.~Nasri,
  %``Neutrino mass textures and partial $\mu-\tau$ symmetry,''
  Phys.\ Rev.\ D {\bf 89}, no. 9, 093004 (2014)
  [\xlink{1311.5869}];
  H.~J.~He and F.~R.~Yin,
  %``Common Origin of $\mu-\tau$ and CP Breaking in Neutrino Seesaw, Baryon 
% Asymmetry, and Hidden Flavor Symmetry,''
  Phys.\ Rev.\ D {\bf 84} (2011) 033009
  [\xlink{1104.2654}];
  S.~F.~Ge, H.~J.~He and F.~R.~Yin,
  %``Common Origin of Soft mu-tau and CP Breaking in Neutrino Seesaw and the Origin 
% of Matter,''
  JCAP {\bf 1005} (2010) 017
  [\xlink{1001.0940}].


\bibitem{hps} L. Wolfenstein, Phys.\ Rev.\ {\bf  D 18}, 958 (1978);
P.~F.~Harrison, D.~H.~Perkins and W.~G.~Scott,
  Phys.\ Lett.\ B {\bf 530}, 167 (2002) [\xlink{hep-ph/0202074}];
Z. z. Xing, Phys.\ Lett.\  {\bf B 533}, 85 (2002).


\bibitem{expt} K. Abe et al.[T2K collaboration], Phys. Rev. Lett. 107 (2011) 041801
[\xlink{1106.2822}]; P. Adamson et al. [MINOS Collaboration], Phys.
Rev. Lett. 107, 181802 (2011) [\xlink{1108.0015}]; Y. Abe et al.
[DOUBLE-CHOOZ Collaboration], Phys. Rev. Lett. 108, 131801 (2012)
[\xlink{1112.6353}]; F. P. An et al. [DAYA-BAY Collaboration], Phys.
Rev. Lett. 108, 171803 (2012) [\xlink{1203.1669}];
J. K. Ahn et al. [RENO Collaboration], Phys. Rev. Lett. 108, 191802 (2012)
[\xlink{1204.0626}].


\bibitem{GL}
P.~F.~Harrison and W.~G.~Scott,
  %``mu - tau reflection symmetry in lepton mixing and neutrino oscillations,''
  Phys.\ Lett.\ B {\bf 547}, 219 (2002)
  [\xlink{hep-ph/0210197}];
W.~Grimus and L.~Lavoura,
  %``A Nonstandard CP transformation leading to maximal atmospheric neutrino
% mixing,''
  Phys.\ Lett.\ B {\bf 579}, 113 (2004)
  [\xlink{hep-ph/0305309}];
% W.~Grimus and L.~Lavoura,
  %``mu-tau Interchange symmetry and lepton mixing,''
  Fortsch.\ Phys.\  {\bf 61}, 535 (2013)
  [\xlink{1207.1678}].

\bibitem{cpmutau:others}
  S.~Gupta, A.~S.~Joshipura and K.~M.~Patel,
  %``Minimal extension of tri-bimaximal mixing and generalized Z_2 X Z_2
% symmetries,''
  Phys.\ Rev.\ D {\bf 85} (2012) 031903
  [\xlink{1112.6113}];
% \bibitem{Ferreira:2012ri}
  P.~M.~Ferreira, W.~Grimus, L.~Lavoura and P.~O.~Ludl,
%   ``Maximal CP Violation in Lepton Mixing from a Model with Delta(27) flavour
% Symmetry,''
  JHEP {\bf 1209} (2012) 128
  [\xlink{1206.7072}].


\bibitem{fcp}
  W.~Grimus and L.~Lavoura, \cite{GL};
%   %``A Nonstandard CP transformation leading to maximal atmospheric neutrino mixing,''
%   Phys.\ Lett.\ B {\bf 579}, 113 (2004)
%   [\xlink{hep-ph/0305309}];
R.~N.~Mohapatra and C.~C.~Nishi,
  %``$S_4$ Flavored CP Symmetry for Neutrinos,''
  Phys.\ Rev.\ D {\bf 86}, 073007 (2012)
  [\xlink{arXiv:1208.2875}];
F.~Feruglio, C.~Hagedorn and R.~Ziegler,
  %``Lepton Mixing Parameters from Discrete and CP Symmetries,''
  JHEP {\bf 1307}, 027 (2013)
  [\xlink{1211.5560}];
M.~Holthausen, M.~Lindner and M.~A.~Schmidt,
  %``CP and Discrete Flavour Symmetries,''
  JHEP {\bf 1304} (2013) 122
  [\xlink{1211.6953}];
%\bibitem{ratz:gcp}
  M.~C.~Chen, M.~Fallbacher, K.~T.~Mahanthappa, M.~Ratz and A.~Trautner,
 % ``CP Violation from Finite Groups,''
  Nucl.\ Phys.\ B {\bf 883} (2014) 267
  [\xlink{1402.0507}];


\bibitem{fcp:others}
L.~L.~Everett, T.~Garon and A.~J.~Stuart,
  %``A Bottom-Up Approach to Lepton Flavor and CP Symmetries,''
  JHEP {\bf 1504}, 069 (2015)
  [\xlink{1501.04336}];
% \bibitem{Chen:2014wxa}
  P.~Chen, C.~C.~Li and G.~J.~Ding,
  %``Lepton Flavor Mixing and CP Symmetry,''
  Phys.\ Rev.\ D {\bf 91} (2015) 3,  033003
  [\xlink{1412.8352}];
% \bibitem{Ding:2014ora}
  G.~J.~Ding, S.~F.~King and T.~Neder,
  %``Generalised CP and $\Delta(6n^2)$ family symmetry in semi-direct models of
% leptons,''
  JHEP {\bf 1412} (2014) 007
  [\xlink{1409.8005}];
% \bibitem{Ding:2014ssa}
  G.~J.~Ding and S.~F.~King,
  %``Generalized $CP$ and $\Delta(96)$ family symmetry,''
  Phys.\ Rev.\ D {\bf 89} (2014) 9,  093020
  [\xlink{1403.5846}];
%\bibitem{ding:a4}
 G.~J.~Ding, S.~F.~King and A.~J.~Stuart,
  %``Generalised CP and $A_4$ Family Symmetry,''
 JHEP {\bf 1312} (2013) 006
 [\xlink{1307.4212}].
% \bibitem{Ding:2013hpa}
  G.~J.~Ding, S.~F.~King, C.~Luhn and A.~J.~Stuart,
  %``Spontaneous CP violation from vacuum alignment in $S_4$ models of leptons,''
  JHEP {\bf 1305} (2013) 084
  [\xlink{1303.6180}];
% \bibitem{Feruglio:2013hia}
  F.~Feruglio, C.~Hagedorn and R.~Ziegler,
  %``A realistic pattern of lepton mixing and masses from $S_4$ and CP,''
  Eur.\ Phys.\ J.\ C {\bf 74} (2014) 2753
  [\xlink{1303.7178}];
% \bibitem{Varzielas:2013eta}
  I.~Medeiros Varzielas and D.~Pidt,
  %``Geometrical CP violation with a complete fermion sector,''
  JHEP {\bf 1311} (2013) 206
  [\xlink{1307.6545}];
% \bibitem{Branco:2015hea}
  G.~C.~Branco, I.~de Medeiros Varzielas and S.~F.~King,
  %``Invariant approach to CP in family symmetry models,''
  \xlink{1502.03105} [hep-ph].
% \bibitem{Ahn:2012tv}
  Y.~H.~Ahn and S.~K.~Kang,
  %``Non-zero $\theta_{13}$ and CP violation in a model with $A_4$ flavor
% symmetry,''
  Phys.\ Rev.\ D {\bf 86} (2012) 093003
  [\xlink{1203.4185}];
% \bibitem{autom}
  C.~C.~Nishi,
  %``Generalized $CP$ symmetries in $\Delta(27)$ flavor models,''
  Phys.\ Rev.\ D {\bf 88} (2013) 3,  033010
  [\xlink{1306.0877}].


\bibitem{babu}
K.~S.~Babu, E.~Ma and J.~W.~F.~Valle,
 % ``Underlying A(4) symmetry for the neutrino mass matrix and the quark mixing
%matrix,''
  Phys.\ Lett.\ B {\bf 552}, 207 (2003)
  [\xlink{hep-ph/0206292}].

\bibitem{cp:real}
E.~Ma,
  %``Transformative A_4 Mixing of Neutrinos with CP Violation,''
  \xlink{1504.02086};
X.~G.~He,
  %``A Model of Neutrino Mass Matrix With $\delta = -\pi/2$ and $\theta_{23} =
% \pi/4$,''
  \xlink{1504.01560}.


\bibitem{deviation}
E.~Ma, A.~Natale and O.~Popov,
  %``Neutrino Mixing and CP Phase Correlations,''
  Phys.\ Lett.\ B {\bf 746}, 114 (2015)
  [\xlink{1502.08023}];
Y.~H.~Ahn, S.~K.~Kang, C.~S.~Kim and T.~P.~Nguyen,
  %``Bridges of Low Energy observables with Leptogenesis in mu-tau Reflection Symmetry,''
  \xlink{0811.1458} [hep-ph].
  
\bibitem{werner} J.~Heeck and W.~Rodejohann,
  %``Gauged $L_mu - L_tau$ Symmetry at the Electroweak Scale,''
  Phys.\ Rev.\ D {\bf 84} (2011) 075007
  [\xlink{1107.5238}];
  J.~Heeck, M.~Holthausen, W.~Rodejohann and Y.~Shimizu,
  %``Higgs ->$\mu\tau$ in Abelian and non-Abelian flavor symmetry models,''
  Nucl.\ Phys.\ B {\bf 896} (2015) 281
[\xlink{1412.3671}].


\bibitem{nuless} W.~Rodejohann,
  %``Neutrino-less Double Beta Decay and Particle Physics,''
  Int.\ J.\ Mod.\ Phys.\ E {\bf 20}, 1833 (2011)
  [\xlink{1106.1334}].


\bibitem{planck}
  P.~A.~R.~Ade {\it et al.}  [Planck Collaboration],
  %``Planck 2015 results. XIII. Cosmological parameters,''
  \xlink{1502.01589} [astro-ph.CO].

\bibitem{nardi}
  S.~Davidson, E.~Nardi and Y.~Nir,
  ``Leptogenesis,''
  Phys.\ Rept.\  {\bf 466} (2008) 105
  [\xlink{0802.2962}]; C.~S.~Fong, E.~Nardi and A.~Riotto,
  %``Leptogenesis in the Universe,''
  Adv.\ High Energy Phys.\  {\bf 2012}, 158303 (2012)
  [\xlink{1301.3062}];
S.~Blanchet and P.~Di Bari,
  %``The minimal scenario of leptogenesis,''
  New J.\ Phys.\  {\bf 14}, 125012 (2012)
  [\xlink{1211.0512}].


\bibitem{flavored}
  A.~Abada, S.~Davidson, F.~X.~Josse-Michaux, M.~Losada and A.~Riotto,
  %``Flavor issues in leptogenesis,''
  JCAP {\bf 0604} (2006) 004
  [\xlink{hep-ph/0601083}];
% \bibitem{Nardi:2006fx}
  E.~Nardi, Y.~Nir, E.~Roulet and J.~Racker,
  %``The importance of flavor in leptogenesis,''
  JHEP {\bf 0601} (2006) 164
  [\xlink{hep-ph/0601084}];


\bibitem{resonant}
  A.~Pilaftsis and T.~E.~J.~Underwood,
  %``Resonant leptogenesis,''
  Nucl.\ Phys.\ B {\bf 692} (2004) 303
  [\xlink{hep-ph/0309342}].

\bibitem{dev}
  P.~S.~Bhupal Dev, P.~Millington, A.~Pilaftsis and D.~Teresi,
  %``Flavour Covariant Transport Equations: an Application to Resonant
% Leptogenesis,''
  Nucl.\ Phys.\ B {\bf 886} (2014) 569
  [\xlink{1404.1003}].

\bibitem{N3:decoupling}
  A.~Ibarra and G.~G.~Ross,
%   ``Neutrino phenomenology: The Case of two right-handed neutrinos,''
  Phys.\ Lett.\ B {\bf 591} (2004) 285
  [\xlink{hep-ph/0312138}].

\bibitem{grimus.ludl}
  W.~Grimus, L.~Lavoura and P.~O.~Ludl,
  %``Is S(4) the horizontal symmetry of tri-bimaximal lepton mixing?,''
  J.\ Phys.\ G {\bf 36} (2009) 115007
  [\xlink{0906.2689}].

\bibitem{gcp}
  H.~Neufeld, W.~Grimus and G.~Ecker,
 % ``Generalized CP invariance, neutral flavor conservation and the structure of
 % the mixing matrix,''
  Int.\ J.\ Mod.\ Phys.\  A {\bf 3} (1988) 603;
%\bibitem{grimus:standard}
  G.~Ecker, W.~Grimus and H.~Neufeld,
  %``A Standard Form For Generalized CP Transformations,''
  J.\ Phys.\ A  {\bf 20} (1987) L807.


\bibitem{hernandez} D.~Hernandez and A.~Y.~Smirnov,
  %``Lepton mixing and discrete symmetries,''
 Phys.\ Rev.\ D {\bf 86}, 053014 (2012)
 [\xlink{1204.0445}];
Phys.\ Rev.\ D {\bf 87}, no. 5, 053005 (2013) [\xlink{1212.2149}].

\bibitem{z2z2}
  S.~F.~Ge, D.~A.~Dicus and W.~W.~Repko,
  %``Z_2 Symmetry Prediction for the Leptonic Dirac CP Phase,''
  Phys.\ Lett.\ B {\bf 702} (2011) 220
  [\xlink{1104.0602}];
%   S.~F.~Ge, D.~A.~Dicus and W.~W.~Repko,
  %``Residual Symmetries for Neutrino Mixing with a Large $\theta_{13}$ and Nearly 
% Maximal $\delta_D$,''
  Phys.\ Rev.\ Lett.\  {\bf 108} (2012) 041801
  [\xlink{1108.0964}].

\bibitem{lam:s4}
  C.~S.~Lam,
  %``Determining Horizontal Symmetry from Neutrino Mixing,''
  Phys.\ Rev.\ Lett.\  {\bf 101} (2008) 121602
  [\xlink{0804.2622}];
%   C.~S.~Lam,
 % ``The Unique Horizontal Symmetry of Leptons,''
  Phys.\ Rev.\ D {\bf 78} (2008) 073015
  [\xlink{0809.1185}].

\bibitem{group-scans}
  C.~S.~Lam,
%  ``Finite Symmetry of Leptonic Mass Matrices,''
  Phys.\ Rev.\ D {\bf 87} (2013) 013001
  [\xlink{1208.5527}];
M.~Holthausen, K.~S.~Lim and M.~Lindner,
 % ``Lepton Mixing Patterns from a Scan of Finite Discrete Groups,''
  Phys.\ Lett.\ B {\bf 721}, 61 (2013)
  [\xlink{1212.2411}];
  M.~Holthausen and K.~S.~Lim,
 % ``Quark and Leptonic Mixing Patterns from the Breakdown of a Common Discrete
%Flavor Symmetry,''
  Phys.\ Rev.\ D {\bf 88} (2013) 033018
  [\xlink{1306.4356}];
%\bibitem{grimus.fonseca}
 R.~M.~Fonseca and W.~Grimus,
 % ``Classification of lepton mixing matrices from finite residual symmetries,''
  JHEP {\bf 1409} (2014) 033
  [\xlink{1405.3678}];
% \bibitem{Talbert:2014bda}
  J.~Talbert,
  %``[Re]constructing Finite Flavour Groups: Horizontal Symmetry Scans from the 
% Bottom-Up,''
  JHEP {\bf 1412} (2014) 058
  [\xlink{1409.7310}].


\bibitem{CMP} 
D.~Chang, R.~N.~Mohapatra and M.~K.~Parida,
  %``Decoupling Parity and SU(2)-R Breaking Scales: A New Approach to Left-Right Symmetric Models,''
  Phys.\ Rev.\ Lett.\  {\bf 52}, 1072 (1984).

\bibitem{seesaw}  P. Minkowski, Phys.\  Lett.\ B {\bf 67}, 421 (1977);
T. Yanagida, Conf. Proc. {\bf C7902131}, 95 (1979);
M. Gell-Mann, P. Ramond and R. Slansky, Conf. Proc. {\bf C790927}, 315 (1979);
S. L. Glashow, NATO Sci. Ser. B {\bf 59}, 687 (1980); R. N. Mohapatra and G. Senjanovi\'{c}, Phys.\ Rev.\  Lett.\  {\bf 44}, 912 (1980);

\bibitem{branco}
  G.~C.~Branco, P.~M.~Ferreira, L.~Lavoura, M.~N.~Rebelo, M.~Sher and J.~P.~Silva,
%  ``Theory and phenomenology of two-Higgs-doublet models,''
  Phys.\ Rept.\  {\bf 516} (2012) 1
  [\xlink{1106.0034}].

\bibitem{2hdm:m}
  H.~E.~Haber and R.~Hempfling,
  %``The Renormalization group improved Higgs sector of the minimal supersymmetric
% model,''
  Phys.\ Rev.\ D {\bf 48} (1993) 4280
  [\xlink{hep-ph/9307201}];
% \bibitem{Pilaftsis:1999qt}
  A.~Pilaftsis and C.~E.~M.~Wagner,
  %``Higgs bosons in the minimal supersymmetric standard model with explicit CP
% violation,''
  Nucl.\ Phys.\ B {\bf 553} (1999) 3
  [\xlink{hep-ph/9902371}].


\bibitem{2hdm:flavor}
  F.~Mahmoudi and O.~Stal,
  %``Flavor constraints on the two-Higgs-doublet model with general Yukawa
% couplings,''
  Phys.\ Rev.\ D {\bf 81} (2010) 035016
  [\xlink{0907.1791}];
% \bibitem{Deschamps:2009rh}
  O.~Deschamps, S.~Descotes-Genon, S.~Monteil, V.~Niess, S.~T'Jampens and
V.~Tisserand,
  %``The Two Higgs Doublet of Type II facing flavour physics data,''
  Phys.\ Rev.\ D {\bf 82} (2010) 073012
  [\xlink{0907.5135}].


\bibitem{pdg}
  K.~A.~Olive {\it et al.}  [Particle Data Group Collaboration],
  %``Review of Particle Physics,''
  Chin.\ Phys.\ C {\bf 38} (2014) 090001.

\bibitem{light.A}
  G.~Abbiendi {\it et al.}  [OPAL Collaboration],
  %``Search for Yukawa production of a light neutral Higgs boson at LEP,''
  Eur.\ Phys.\ J.\ C {\bf 23} (2002) 397
  [\xlink{hep-ex/0111010}];
  J.~Abdallah {\it et al.}  [DELPHI Collaboration],
  %``Searches for neutral higgs bosons in extended models,''
  Eur.\ Phys.\ J.\ C {\bf 38} (2004) 1
  [\xlink{hep-ex/0410017}].
% \bibitem{Dolan:2014ska}
  M.~J.~Dolan, C.~McCabe, F.~Kahlhoefer and K.~Schmidt-Hoberg,
  %``A taste of dark matter: Flavour constraints on pseudoscalar mediators,''
  JHEP {\bf 1503} (2015) 171
  [\xlink{1412.5174}].


\bibitem{craig}
For a review and earlier references, see N.~Craig, F.~D'Eramo, P.~Draper, S.~Thomas 
and H.~Zhang,
  %``The Hunt for the Rest of the Higgs Bosons,''
  \xlink{1504.04630} [hep-ph].


 \bibitem{nir}
  A.~Efrati and Y.~Nir,
  %``What if $\lambda_{hhh}\neq 3m_h^2/v$,''
  \xlink{1401.0935} [hep-ph].


\end{thebibliography}
\end{document}